\documentclass[%
 reprint,
superscriptaddress,
nofootinbib,
 amsmath,amssymb,
 aps,
]{revtex4-2}
\usepackage{color}
\usepackage{graphicx}
\usepackage{dcolumn}
\usepackage{bm}
\usepackage{hyperref}

\begin{document}

\preprint{APS/123-QED}

\title{Time-dependent SI model for epidemiology and applications to Covid-19}

\author{L. Arturo Ure\~{n}a-L\'{o}pez} 
 \email{lurena@ugto.mx}
\affiliation{%
Departamento de F\'isica, DCI, Campus Le\'on, Universidad de
Guanajuato, 37150, Le\'on, Guanajuato, M\'exico.}

\author{Alma X. Gonz\'alez-Morales}%
 \email{alma.gonzalez@fisica.ugto.mx}
 \affiliation{%
Departamento de F\'isica, DCI, Campus Le\'on, Universidad de
Guanajuato, 37150, Le\'on, Guanajuato, M\'exico.}
 \affiliation{Consejo Nacional de Ciencia y Tecnolog\'ia,
Av. Insurgentes Sur 1582. Colonia Cr\'edito Constructor, Del. Benito   Ju\'arez C.P. 03940, M\'exico D.F. M\'exico}

\date{\today}

\begin{abstract}
A generalisation of the Susceptible-Infectious model is made to include a time-dependent transmission rate, which leads to a close analytical expression in terms of a logistic function. The solution can be applied to any continuous function chosen to describe the evolution of the transmission rate with time. Taking inspiration from real data of the Covid-19, for the case of cumulative confirmed positives and deaths, we propose an exponentially decaying transmission rate with two free parameters, one for its initial amplitude and another one for its decaying rate. The resultant time-dependent SI model, which under extra conditions recovers the standard Gompertz functional form, is then compared with data from selected countries and its parameters fit using Bayesian inference. We make predictions about the asymptotic number of confirmed positives and deaths, and discuss the possible evolution of the disease in each country in terms of our parametrisation of the transmission rate.
\end{abstract}

\maketitle

\section{Introduction}
The epidemics of Covid-19 has prompted a lot of researchers from different fields to use their diverse expertise to understand the nature of the Sars-Cov-2 virus and its spreading worldwide. In particular, out of the medical sciences, the epidemics has revealed a rich ground where physicists, mathematicians and statisticians are eager to apply their knowledge and contribute to its ameriolation and containment at the local and global stages.

From the formal point of view, there has been a lot of research in the so-called epidemic models and their mathematical properties, see for instance~\cite{Brauer2010,ahmada2020number}, being the compartmental models the most widely used and studied. The models in general consider the classification of a given population in some parts: Susceptible (S), Infective (I), Recovered (R), among others, and are then dubbed in terms of which of them they consider for the dynamics of the disease: SIR, SEIR, etc. There is historical evidence that such models are in good agreement with the dynamics of past epidemics
~\cite{Brauer2010}, and such past successes have triggered its use in the present crisis, see for instance~\cite{arm2020predicting,afonso2020epidemic,hult2020estimates,piasecki2020new,far2020modelling,khan2020predictive,pulido2020geospatial,capistran2020forecasting,barrio2020modelling}. 

However, there is an inherent difficulty in studying the Covid-19 epidemics in real time: that data collection is not perfect, and in most cases is certainly incomplete and not very useful to fully characterize the evolution of the epidemics~\cite{zelner2020accounting,altmejd2020nowcasting}. One then must question whether the use of very complex models is convenient, given the scarcity and flaws of the data series. Actually, it seems that simple models are sufficient to understand the epidemics~\cite{carletti2020covid19,fiacchini2020ockhams}.

Given the above considerations, here we study the simplest of compartmental models and use it for the description of the current epidemics in different countries. The model only accounts for two parts of the population: Susceptibles and Infectives, and is known as the SI model. The infectives in the model are not supposed to recover, and then its number is an ever increasing function of time, which is what one sees in the daily reports of cumulative infectives released by health authorities worldwide. In this respect, the SI model seems to be a convenient one to follow the evolution of the real data at hand.

However, the original SI model has one free parameter, the transmission rate, which is assumed to be a constant parameter, but this seems to be an oversimplification that is not in agreement with real data. This has inspired the use of time-dependent transmission rates which expand the capabilities of the compartmental models to encompass hidden complexities of the data compilations. Here we take this point of view and consider a time-dependent transmission rate in terms of an exponentially decreasing function, as it was done first in
~\cite{bizet2020modelos}, and also in~\cite{palladino2020modelling} (see also~\cite{chao2020simplified,nesterov2020online} for other examples). Additionally, we also take into account the data series of cumulative deaths, by means of simple assumptions withouth extending the SI model, which are thought to be at least as reliable of those of cumulative infectives, and probably a better representation of the epidemics course~\cite{carletti2020covid19}.

A summary of the paper is as follows. In Sec.~\ref{sec:mathematical}, we present the mathematical description of the standard SI model, which we call the time-independent version and give a brief account of its main properties. We then explain its generalization to accommodate a time-dependent transmission rate and find the analytical solution in terms of a logistic function for the general case. It is argued that, because of the time-dependence of the transmission rate, the total population number can be freely choose and there is a simpler, approximated solution of the time-dependent SI system in the limit of a large population number in the form of a Gompertz function~\cite{gompertz,Winsor1932}. Using real data, we look for evidence of a time-dependent transmission rate in terms of its definition within the SI system, and then we propose an exponential-like parametrisation of the transmission rate. 

In Sec.~\ref{sec:statistical} we use the time-dependent models of Sec.~\ref{sec:mathematical} and use them to make a fitting of their free parameters to data from some chosen countries. The fit is made by means of the Bayesian inference, and the results are giving an interpretation in terms of characteristic quantities and times that are intrinsic to our parametrised transmission rate. Finally, the main conclusions of our study and future perspectives are summarized in Sec.~\ref{sec:final}.

\section{Mathematical background \label{sec:mathematical}}
Here we present the main mathematical expressions for the time-dependent epidemic model, based on the known compartmental model SI.

\subsection{Time-independent SI \label{sec:time-independent}}
The SI model is represented by the following set of equations,
\begin{equation}
    \dot{S} = - \beta S \frac{I}{N} \, , \quad \dot{I} = \beta S \frac{I}{N} \, , \label{eq:si}
\end{equation}
Here, $S$ ($I$) is the number of susceptible (infectious) people, the total population is $N=S+I$, $\beta$ is the infection rate (ie the probability per unit time that an individual contracts the disease), and a dot means derivative with respect to time.\footnote{It is possible to define new normalized variables in the form $\hat{S} = \beta S/N$ and $\hat{I} = \beta I/N$, and then Eqs.~\eqref{eq:si} are simply written as $\dot{\hat{S}} = - \hat{S} \hat{I}$ and $\dot{\hat{I}} = \hat{S} \hat{I}$. The constraint equation also becomes $\hat{S} + \hat{I} = \beta$, and then the transmission rate only appears for the initial conditions.} 

Because of the conservation equation, the SI system is truly one-dimensional and represented by the equation,
\begin{equation}
    \dot{I} = \beta \left(1 - I/N \right) I \, . \label{eq:si-onedim}
\end{equation}
If $\beta$ is a constant parameter, the solution of Eq.~\eqref{eq:si-onedim} is the sigmoid, also known as the logistic function~\cite{verhulst},
\begin{equation}
    I = \frac{N}{1+e^{\beta (t_0-t)}} \, , \label{eq:logistic}
\end{equation}
where $t_0$ is an integration constant. 

From the second derivative of Eq.~\eqref{eq:logistic} (the first derivative of Eq.~\eqref{eq:si-onedim}), 
\begin{equation}
    \ddot{I} = \beta \dot{I} \left( 1 - 2 I/N \right) \, . 
\end{equation}{}
Then, there is a maximum of the first derivative at $t=t_0$, that also corresponds to an inflection point of the logistic function ($\ddot{I} =0$) at which $I=N/2$. As for the logistic function itself, notice that the initial and asymptotic values of the logistic function~\eqref{eq:logistic} are, respectively,
\begin{equation}
    I_i = \frac{N}{1+e^{\beta t_0}} \, , \quad I_\infty = N \, . \label{eq:si-constants}
\end{equation}
In other words, the saturation value of the infectious people is the whole of the available population. If $N$ is a very large number, and for the same $I_i$, the only change is the position of the inflection point $t_0$, which just shifts to larger values.

\subsection{Time-dependent SI system \label{sec:time-dependent}}
Let us now consider the case of a time-dependent infection rate, that is, $\beta = \beta(t)$. Notice that we do not need to change the nature of the original SI system~\eqref{eq:si}, and then we can use again Eq.~\eqref{eq:si-onedim} to find a solution of a time-dependent SI system. It can be readily shown that the solution can be written as a generalized logistic function~\cite{pearlreed,Winsor1932} (see also~\cite{Tsoularis2002}),
\begin{subequations}
\label{eq:si-td}
\begin{equation}
     I(t) = \frac{N}{1+e^{u_0-u}} \, , \quad \mathrm{with} \quad u(t) = \int^t_0 \beta(x) \, dx \, . \label{eq:logistic-u}
\end{equation}
where $u_0$ is an integration constant. Notice that Eq.~\eqref{eq:logistic-u} is again the sigmoid function but only now in terms of a new variable $u$. The initial and asymptotic values of the infected people are given by
\begin{equation}
    I_i = \frac{N}{1+e^{u_0}} \, , \quad I_\infty = \frac{N}{1+e^{u_0-u_\infty}} \, . \label{eq:si-constants2}
\end{equation}
Here, $u_\infty = u(t \to \infty)$, which may or may not be a finite value, and this depends on the chosen function $\beta$, see Eq.~\eqref{eq:logistic-u}.

Notice that the second derivative of Eq.~\eqref{eq:si-onedim} for a time-dependent transmission rate is
\begin{equation}
    \ddot{I} = \dot{I} \left[ \frac{\dot{\beta}}{\beta} + \beta \left( 1 - 2I/N \right) \right] \, . \label{eq:Sderivative}
\end{equation}
\end{subequations}
Hence, the true inflection point $ \ddot{I} =0$ does not correspond to $I=N/2$ anymore as it was in the time-independent case. However, it can be shown that the time-independent SI solution~\eqref{eq:logistic} is a particular case of the time-dependent one~\eqref{eq:logistic-u}. In the case $\beta = \mathrm{const.}$, one readily obtains that $u(t)= \beta t$, and then Eq.~\eqref{eq:logistic} is recovered if also $u_0 = \beta t_0$.

The exact solution~\eqref{eq:logistic-u} opens up the possibility to consider the evolution of a disease in which the transmission rate is changing, whether by natural means or by human intervention. This is a more realistic approach, and it has the advantage that we can continue dealing with the accumulated numbers reported by the health systems worldwide. 

\subsection{The large $N$ limit}
\label{sec:large-N}
In the time-independent SI system, the constant of integration is fixed by the initial conditions and the total population number, namely $e^{\beta t_0}= N/I_i-1$, and then the asymptotic value of the infectives is the total population, see Eqs.~\eqref{eq:si-constants}. However, in the time-dependent case the asymptotic value of infectives also depends on the asymptotic value $u_\infty$, which means that not all the population will get infected, that is $I_\infty < N$. 

The ratio between the total and the initial number of infectives can be written as
\begin{equation}
    \frac{I_\infty}{I_i} = \frac{1+e^{u_0}}{1+C e^{u_0-u_\infty}} \, . \label{eq:Iratios}
\end{equation}
Here, $u_\infty$ is an independent constant, and then one can obtain the same ratio by adjusting accordingly the values of $u_0$ and $u_\infty$. In this sense, there is a new degeneracy in the time-dependent system as the total number of infectives is not uniquely determined by the total population number $N$. Explicitly, the degeneracy reads
\begin{equation}
    e^{u_\infty} = \frac{(I_\infty/I_i) e^{u_0}}{1+e^{u_0}- (I_\infty/I_i)} \, . \label{eq:degeneracy}
\end{equation}
If we keep the ratio $(I_\infty /I_i)$ fixed, we obtain that
\begin{equation}
    \lim_{u_0 \to \infty} e^{u_\infty} = \frac{I_\infty}{I_i} \, . \label{eq:degeneracy-limit}
\end{equation}
The final ratio of infectives, in this limit, can be calculated directly from the asymptotic value of the variable $u$. Given that $e^{u_0}= N/I_i-1$, we call this the large $N$ limit.

Actually, one can do the same exercise in the general expression of infectives. If we write Eq.~\eqref{eq:logistic-u} (see also Eq.~\eqref{eq:si-constants2}) in the form,
\begin{equation}
    I(t) = \frac{1+e^{u_0}}{e^u + e^{u_0}} I_i e^u \, , \label{eq:degeneracy1}
\end{equation}
we find that for large enough values of $u_0$, ie $N \to \infty$, the function of infectives can be approximated as
\begin{subequations}
\begin{equation}
    I(t) \simeq I_i e^u = I_i \exp \left[ \int^t_0 \beta(x) \, dx \right] \, . \label{eq:degeneracy1-limit}
\end{equation}
The evolution of the disease is simply driven by the transmission rate, and then the asymptotic limit is obtained if the integral on the rhs of~\eqref{eq:degeneracy1-limit} converges for $t \to \infty$. Actually, the ratio between the asymptotic and initial values of infectives is
\begin{equation}
    \frac{I_\infty}{I_i} \simeq e^{u_\infty} = \exp \left[ \int^\infty_0 \beta(x) \, dx \right] \, , \label{eq:degeneracy2-limit}
\end{equation}
\end{subequations}
which is in turn the same result as in Eq.~\eqref{eq:degeneracy-limit} above.

The expression~\eqref{eq:degeneracy1-limit} gives a simplified evolution of the disease. For instance, the first derivative is just $\dot{I} = \beta I$, whereas the second derivative is $\ddot{I} = (\dot{\beta} + \beta^2) I$, and then the inflection point $t_0$ in the evolution curve of infectives is given by the solution of the equation $(\dot{\beta} + \beta^2 )(t_0) =0$.

There is a small warning here about the parametrisation of $\beta$. If we assume that the transmission rate is given by $\beta=\beta(t,k_\ell)$, where $k_\ell$ are in general constant parameters, one must be aware that their values will depend on $N$, and then the values obtained from Eq.~\eqref{eq:degeneracy1-limit} are not the same as those from Eq.~\eqref{eq:degeneracy1}. They will, in both cases, deliver the same values of $I_i$ and $I_\infty$, but the evolution profile $I(t)$ will have some differences in the two cases. 

\subsection{Time-dependent $\beta$ from real data \label{sec:time-dependent-data}}
One important question is whether real data suggests a complex evolution of the Covid-19 disease, for which the simple time-independent SI system would be insufficient to describe it. To get an answer directly from the data available, we make use here of an analytical result that can be derived from the general case~\eqref{eq:logistic-u}. 

First, we write down the following expression to define the time-dependent function $\Gamma (t)$,
\begin{subequations}
\label{eq:limit-k1}
\begin{equation}
    \Gamma(t) \equiv \frac{\dot{I}}{I(1-I/I_\infty)} = \frac{\beta (t)}{1 - e^{u - u_\infty}} \, , \label{eq:limit-k1a}
\end{equation}
which is valid for any form of $\beta$ and is directly derived from Eqs.~\eqref{eq:si-td}; it is also valid for the large $N$ approximation~\eqref{eq:degeneracy1-limit}. In the time-independent case, for which $I_\infty = N$, we readily obtain $\Gamma(t) =\beta = \mathrm{const.}$ But in the general case we find at $t=0$ that
\begin{equation}
    \Gamma (0) = \frac{\beta(0)}{1 - e^{- u_\infty}} \, . \label{eq:limit-k1b}
\end{equation}

Another characteristic value is found at late times, as $u \to u_\infty$. In this case, we use the approximation $1-e^{u - u_\infty} \simeq u - u_\infty$, and then from Eq.~\eqref{eq:logistic-u} we obtain that
\begin{equation}
    \lim_{t \to \infty} \Gamma (t) = - \lim_{t \to \infty} \left[ \frac{1}{\beta(t)} \int^\infty_t \beta(x) dx \right]^{-1} \, , \label{eq:limit-k1c}
\end{equation}
\end{subequations}
if such limit exists. 

\begin{figure*}[htp!]
\includegraphics[width=0.49\textwidth]{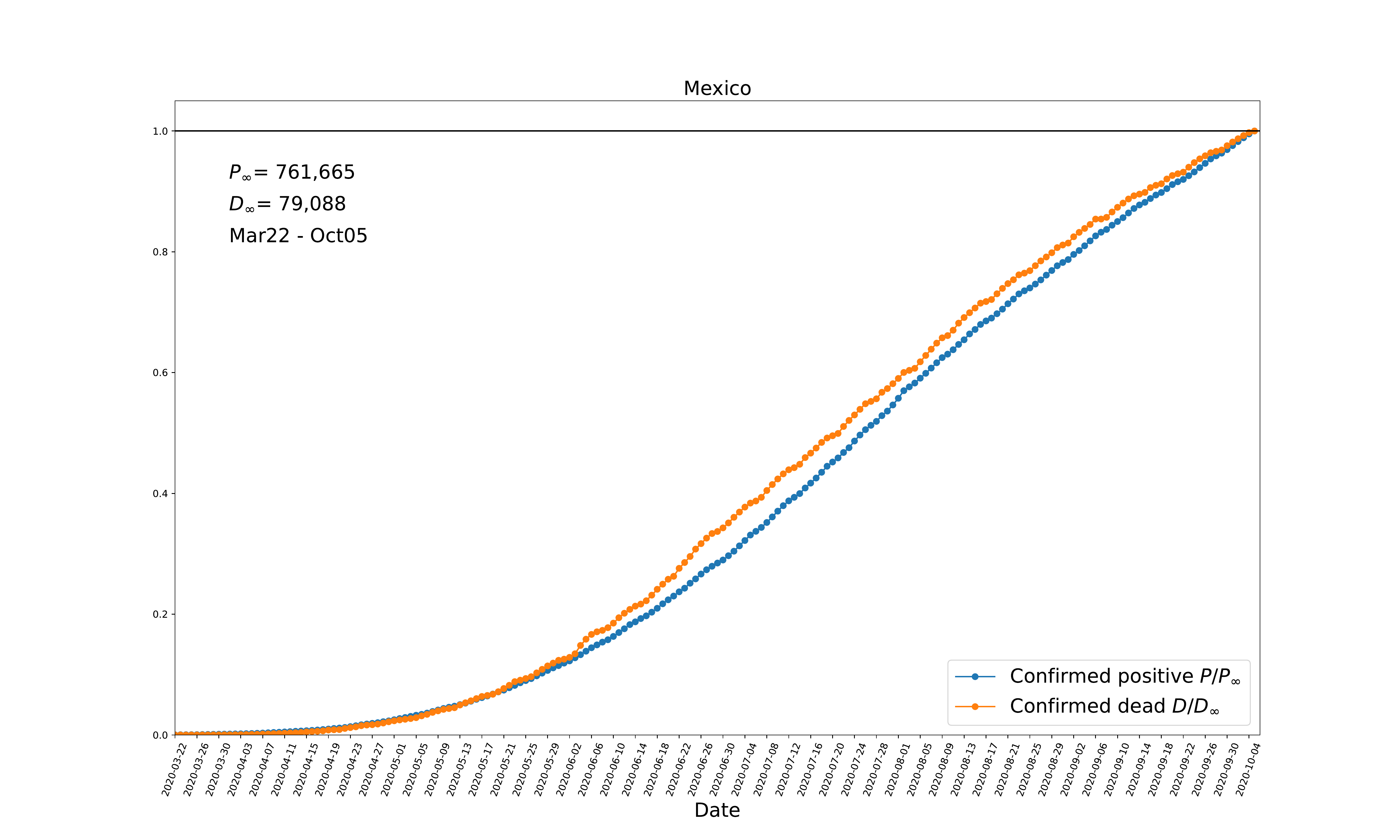}
\includegraphics[width=0.49\textwidth]{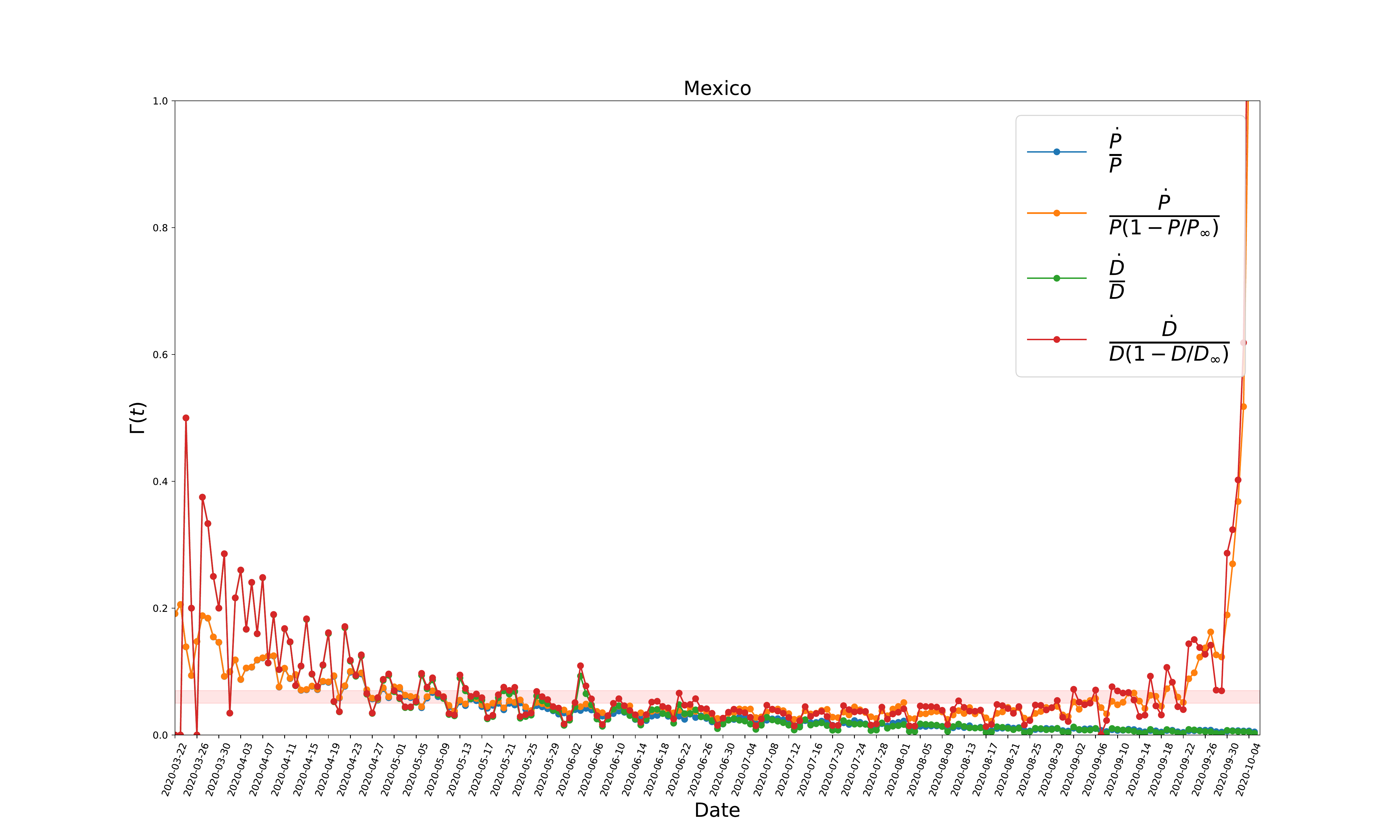}
\caption{\label{fig:beta} (Left) Data from Mexico showing the evolution of accumulated confirmed positives and deaths, the time series start at the date of the first registered death (from March 22 until October 5). The values in the plot are normalized with their respective last data point. It can be noticed that both quantities seem to have reached an asymptotic final value. (Right) The estimated evolution of function $\Gamma(t)$, see Eq.~\eqref{eq:limit-k1a}, according to data of new daily cases for both confirmed positives and deaths. Shown are three different combinations of each data compilation, where $P_\infty$ and $D_\infty$ are the last data points in the corresponding time series. The trends of the three cases suggest a decreasing of the transmission rate $\beta$ with time. The orange-shaded horizontal region (with vertical range $0.05-0.07$) is just shown for reference. See Sec.~\ref{sec:time-dependent-data} and the text for more details.}
\end{figure*}

Thus, the function $\Gamma(t)$ can help us to decide whether to use the time-independent or the time-dependent SI system, as the expression~\eqref{eq:limit-k1a} can be easily calculated from data in countries that already show an asymptotic value of infectives. In Fig.~\ref{fig:beta} we show data from Mexico\footnote{All data considered in this work have been taken from: \texttt{https://ourworldindata.org/coronavirus-source-data}}, which is one of the countries that comply with the foregoing condition, as seen on the left panel for the number of positives normalized to the latest reported value (the first day in the time series is that of the first registered death). Shown also is the evolution of reported deaths, which is too normalized to the value of the latest reported value.

On the right panel of Fig.~\ref{fig:beta} we see the daily new cases (with the same aforementioned normalization) divided by different combinations of accumulated numbers. In the plots $P$ ($D$) refers to confirmed positives (deaths). Notice that the combinations in the plot represent the function~\eqref{eq:limit-k1a}, and then data seems to indicate that $\Gamma(t)$ is a decreasing function that approaches a constant value at late times. In contrast, when we use the expression $\dot{I}/I$, the ratio goes to zero. Our interpretation is that data suggest a time-dependent transmission rate $\beta$, particularly one that decays with time. Although we only show the case of Mexico, we have found the same trend in all cases in which data already shows an asymptotic value of accumulated positives (eg, Germany, France, and others).

It is clearly seen that $\beta$ is in general a decaying function as time proceeds, which in our opinion is a manifestation of governmental intervention to slow the spread of the disease. Although there is not a unique function, given the profile suggested in Fig.~\ref{fig:beta}, we will consider an exponentially decaying function as an approximation to the evolution of $\beta$. That is, we write it explicitly as
\begin{subequations}
\label{eq:betatd}
\begin{equation}
    \beta(t) = k_0 e^{-k_1 t} \, , \label{eq:betatd-a}
\end{equation}
where $k_0,k_1$ are constant parameters. The corresponding time variable in Eq.~\eqref{eq:logistic-u} is
\begin{equation}
    u(t) = \frac{k_0}{k_1} \left( 1 - e^{-k_1 t} \right) \, , \label{eq:betatd-b}
\end{equation}
\end{subequations} 
and then for this case we see that $u_\infty = k_0/k_1$. 

In the limit $k_1 t \ll 1$ we obtain that $u(t) \simeq k_0 t$, which is the expected behavior in the time-independent case. Then, $k_0$ determines the initial growth of the epidemics, whereas $k_1$ gives the decay time of the transmission rate, with a half-life time given by $t_{1/2} = \ln 2/k_1$.

Likewise, the asymptotic value of the infectives is $I_\infty = I_i e^{k_0/k_1}$, see Eqs.~\eqref{eq:degeneracy-limit} and~\eqref{eq:degeneracy2-limit}, and then the ratio between the asymptotic and initial values will depend on the ratio $k_0/k_1$. Here we see the importance of $k_1$: the smaller its value, the larger it takes for the epidemics to fade away and the larger the asymptotic value of total infectives.

As for the time-dependent function $\Gamma(t)$ defined in Eq.~\eqref{eq:limit-k1a}, after using Eqs.~\eqref{eq:betatd} we find that
\begin{eqnarray}
    \Gamma(0) = \frac{k_0}{1 - e^{-k_0/k_1}} \, , \quad \lim_{t \to \infty} \Gamma (t) = k_1 \, . \label{eq:limit-k1d}
\end{eqnarray}
Notice that Eq.~\eqref{eq:limit-k1d} is also in agreement with the expectation from data in Fig.~\ref{fig:beta}: there are well definite values of $\Gamma(t)$ at early and late times at late times, where the value at late times will give us an indication of decay time of the transmission rate $\beta$.

We mentioned before that it is not possible to have an analytical expression for the inflection time of the sigmoid function~\eqref{eq:logistic-u} for a general $\beta(t)$. However, one can instead write an expression for the time at which the infected population $I(t_{50})$ is one half ($50\%$) of the asymptotic value, ie $I(t_{50}) =I_\infty /2$. After some straightforward algebra using Eqs.~\eqref{eq:si-td}, we find in general for the generalized time variable that
\begin{subequations}
\label{eq:crossing}
\begin{equation}
   u_{50} = u_0 - \ln \left( 1+ 2 e^{u_0 - u_\infty} \right) \, . \label{eq:crossing-a}
\end{equation}
Hence, from the particular parametrisation~\eqref{eq:betatd-a} we finally obtain
\begin{equation}
    t_{50} = \frac{1}{k_1} \ln \left( \frac{k_0/k_1}{k_0/k_1 - u_{50}} \right) \, . \label{eq:crossing-b}
\end{equation}
\end{subequations}

We will use $t_{50}$ as a point of reference in the time series of the different data in our analysis below, similar to the inflection point of the standard sigmoid function. Another useful point is that at which the number of infectives is $90\%$ of the asymptotical value, $I(t_{90}) =0.9 I_\infty$, as it can be considered as a reference for the upcoming end of the infection period. Following the same calculations that led to Eqs.~\eqref{eq:crossing}, we find
\begin{subequations}
\label{eq:crossing90}
\begin{equation}
   u_{90} = u_0 - \ln \left[ 1/9+ (10/9) e^{u_0 - u_\infty} \right] \, , \label{eq:crossing90-a}
\end{equation}
and its corresponding time, again for the particular parametrisation~\eqref{eq:betatd-a}, is
\begin{equation}
    t_{90} = \frac{1}{k_1} \ln \left( \frac{k_0/k_1}{k_0/k_1 - u_{90}} \right) \, . \label{eq:crossing90-b}
\end{equation}
\end{subequations}

\section{Statistical analysis and results \label{sec:statistical}}
Our main assumption is that the time series of both confirmed positives and deaths have a common origin from the total number of infected people $I(t)$. Formally speaking, we are assuming that
\begin{equation}
    P(t) = r_P I(t) \, , \quad D(t) = r_D I(t-t_D) \, , \label{eq:infection}
\end{equation}
where $r_P$ is the fraction of infected people that is tested and confirmed as positive, whereas $r_D$ is the fraction of infected people that is confirmed positive and eventually pass away. Notice that we consider that $D(t)$ evolve with a time delay $t_D$ with respect to $I(t)$ (and also to $P(t)$). Such delay is difficult to measure reliably and here we will report the values suggested by the data itself.

In the following sections we do the fitting to data using two different models. The first one, which we call model A, considers the generalized logistic function~\eqref{eq:logistic-u} and the parametrisation~\eqref{eq:betatd-a} of the transmission rate. The second one, which we dub model B, uses the large $N$ approximation represented by Eq.~\eqref{eq:degeneracy1-limit} and the same parametrisation of the transmission rate. As discussed previously in Sec.~\ref{sec:time-dependent-data}, data seems to discard the time-independent SI system, but for completeness we also report its fitting to data in Appendix~\ref{sec:modelC}.

\subsection{Model A and fitting to data \label{sec:modelA}}
We will consider the following parametrization of the logistic function of the confirmed positive people, in cumulative numbers,
\begin{subequations}
\label{eq:modela}
\begin{equation}
     P(t) = P_i \, \frac{1+e^{u_0}}{1+e^{u_0-u(t)}} \, . \label{eq:positive-a}
\end{equation}
Following the discussion at the beginning of this section, we will also consider the reports of accumulated deaths, which we assume to follow a similar logistic function as those of the confirmed positive, except for a different amplitude and inflection point,
\begin{equation}
     D(t) = D_i \, \frac{1+e^{\hat{u}_0}}{1+e^{\hat{u}_0-u(t-t_D)}} \, , \label{eq:deaths-a}
\end{equation}
\end{subequations}
Our model then has seven free parameters: $P_0$, $D_0$, $t_D$, $u_0$, $\hat{u}_0$, $k_0$ and $k_1$, and then one does not hope for tight constraints on them given the scarcity of data about Covid-19 infections in general. Additionally, one must remember that data is not generated in a systematic way as in a laboratory experiment, and there may be many sources of error in the data management and processing of new cases.

We will assume that the data provided follows the trend of the real number of infected people and deaths, and that both of these numbers will eventually reach a saturation value in the near future, as it has happened in past diseases. Moreover, as we do not know the systematic errors in the processing of the data, we will assume some level of intrinsic error by using a Poisson distribution. To ease the fitting of data from different countries we normalize the data so that the first point in the time series is of order of unity. Given this, we consider flat priors in the form: $0 < P_i < 10$, $0< D_i < 10$, $0 < t_D < 100$, $0 < u_0, \, \hat{u}_0 < 30$, and $0< k_0, \, k_1 < 3$.

\begin{figure*}[htp!]
\includegraphics[width=\textwidth]{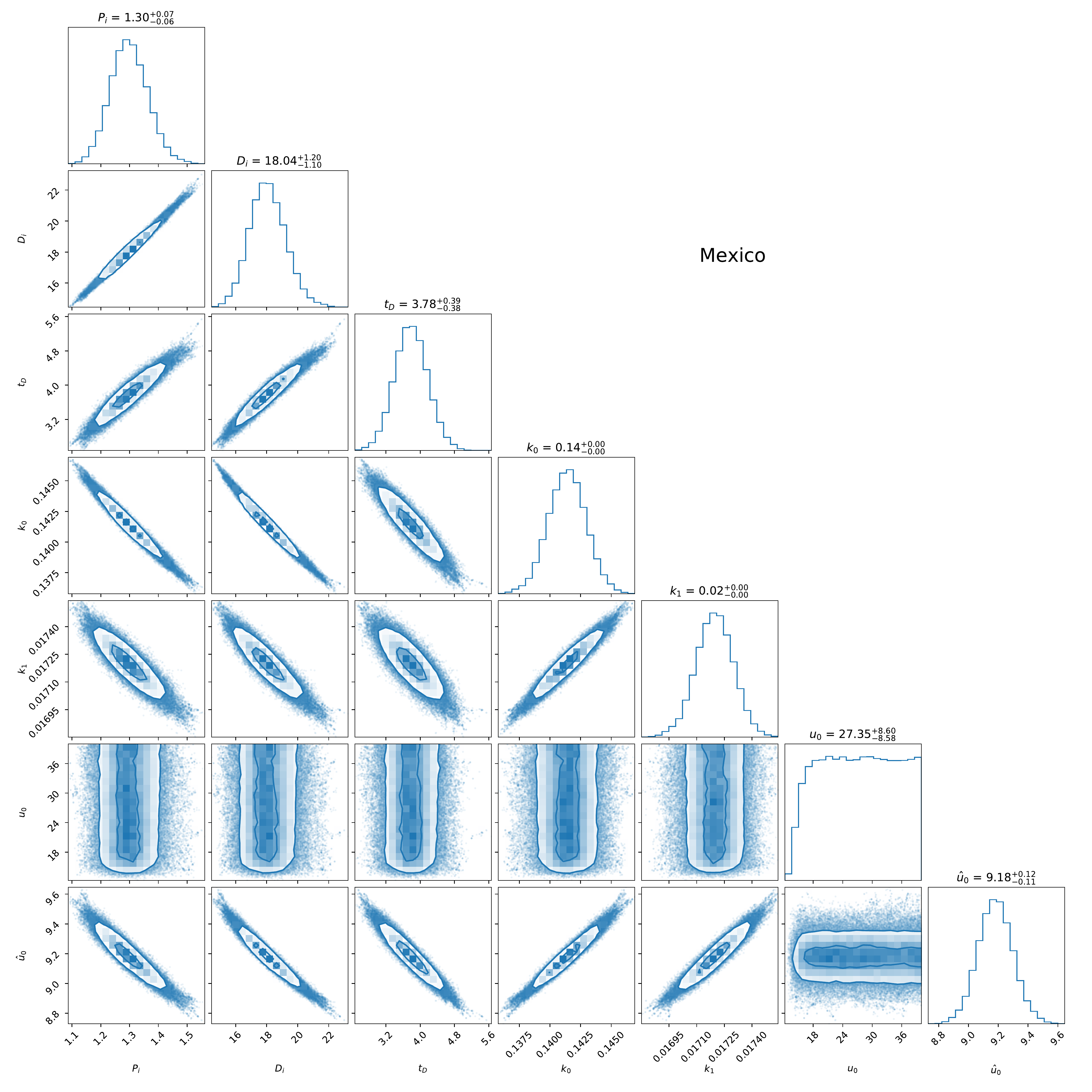}
\caption{\label{fig:modela-a} Triangle plot of the fitting to data from Mexico (see also Fig.~\ref{fig:beta}) of the parameters $P_i$, $D_i$, $t_D$, , $k_0$, $k_1$, $u_0$ and $\hat{u}_0$ of model A described in Sec.~\ref{sec:modelA}. In general, all parameters are well constrained by the combination of accumulated confirmed positives and deaths. See the text for more details.}
\end{figure*}

To fit the data, we will use the EMCEE algorithm by means of a Python script, using 100 chains with 30000 steps in each one. The results are shown in Fig.~\ref{fig:modela-a} for the free parameters of our model. A common feature of all the countries we considered, not just the one shown in Fig.~\ref{fig:modela-a}, is that the values of $P_i$, $D_i$, $t_0$, $k_0$, $k_1$, $u_0$ and $\hat{u}_0$, are all well constrained by the data, which is consistent with our assumption that both confirmed positives and deaths follow the same trend of evolution. 

We have applied our model to the data of 9 other countries, which at the moment of writing are the countries with the highest number in cumulative positives and deaths in terms of their population size, and the fitting results are summarized in Table~\ref{tab:1a}. As said before, $P_\infty$ and $D_\infty$ are the expected final numbers for the cumulative positives and deaths in each case, even for countries that have not yet reached an asymptotic value.

The delay time between positives and deaths, $t_D$, is for all cases lower than 20 days, which is consistent with the general fact that all deaths were first confirmed as positive, and then $t_D$ will represent the delay time between a positive test and the occurrence of death. 

Next are the parameters of the transmission rate $\beta$, where we note a strong similarity in the values of the different countries. First, we recall that $\beta(0)=k_0$, and then this parameter is the value of the transmission rate at the start of the time series. Likewise, parameter $k_1$ is the decay rate of the transmission rate. The last two columns in Table~\ref{tab:1a} show the values of the integration constants $u_0$ and $\hat{u}_0$, which are directly related to the total population number $N$. 

\begin{table*}[htp!]
\caption{\label{tab:1a} Fitted values of parameters in model A, see Eqs.~\eqref{eq:modela}, as obtained from the data of different countries. The confidence regions for the parameters in the case of Mexico are shown in Fig.~\ref{fig:modela-a}.}
\begin{ruledtabular}
\begin{tabular}{|c|c|c|c|c|c|c|c|}
Country & $P_\infty$ & $D_\infty$ & $t_D$ & $k_0$ & $k_1$ (1/days) & $u_0$ & $\hat{u}_0$ \\ 
\hline
Mexico & $980,256 \pm^{17,593}_{16,645}$ & $97,084 \pm^{487}_{484}$ & $3.78 \pm^{1.02}_{0.98}$ & $0.14 \pm^{0.00}_{0.00}$ & $0.017 \pm^{0.000}_{0.000}$ & $27.35 \pm^{12.52}_{13.50}$ & $9.18 \pm^{0.31}_{0.30}$ \\
\hline
Peru & $1,075,559 \pm^{13,086}_{12,868}$ & $51,066 \pm^{537}_{524}$ & $12.00 \pm^{0.58}_{0.58}$ & $0.10 \pm^{0.00}_{0.00}$ & $0.015 \pm^{0.000}_{0.000}$ & $26.19 \pm^{13.67}_{14.75}$ & $26.19 \pm^{13.66}_{14.83}$ \\
\hline
Belgium & $77,073 \pm^{1,464}_{3,321}$ & $9,802 \pm^{23}_{22}$ & $9.82 \pm^{0.27}_{8.00}$ & $0.30 \pm^{0.10}_{0.01}$ & $0.046 \pm^{0.004}_{0.009}$ & $6.47 \pm^{0.93}_{0.29}$ & $4.78 \pm^{2.16}_{0.06}$ \\
\hline
Bolivia & $187,445 \pm^{7,699}_{6,992}$ & $15,807 \pm^{1,547}_{1,323}$ & $0.06 \pm^{0.42}_{0.06}$ & $0.09 \pm^{0.01}_{0.01}$ & $0.009 \pm^{0.001}_{0.001}$ & $7.13 \pm^{0.24}_{0.21}$ & $8.27 \pm^{0.38}_{0.30}$ \\
\hline
Brazil & $6,968,037 \pm^{50,851}_{51,706}$ & $166,676 \pm^{356}_{362}$ & $0.00 \pm^{0.01}_{0.00}$ & $0.13 \pm^{0.00}_{0.00}$ & $0.016 \pm^{0.000}_{0.000}$ & $28.77 \pm^{11.12}_{12.35}$ & $8.03 \pm^{0.03}_{0.02}$ \\
\hline
Chile & $442,318 \pm^{7,888}_{14,407}$ & $12,488 \pm^{98}_{87}$ & $19.47 \pm^{4.32}_{18.54}$ & $0.09 \pm^{0.01}_{0.00}$ & $0.006 \pm^{0.000}_{0.000}$ & $6.19 \pm^{0.56}_{0.13}$ & $5.87 \pm^{1.71}_{0.37}$ \\
\hline
Ecuador & $138,598 \pm^{1,108}_{1,101}$ & $10,568 \pm^{61}_{62}$ & $0.01 \pm^{0.07}_{0.01}$ & $0.11 \pm^{0.00}_{0.00}$ & $0.019 \pm^{0.000}_{0.000}$ & $8.46 \pm^{0.39}_{0.27}$ & $26.70 \pm^{13.17}_{14.24}$ \\
\hline
United States & $6,559,124 \pm^{9,639}_{9,778}$ & $182,774 \pm^{141}_{140}$ & $0.00 \pm^{0.00}_{0.00}$ & $0.19 \pm^{0.00}_{0.00}$ & $0.022 \pm^{0.000}_{0.000}$ & $29.97 \pm^{9.93}_{10.72}$ & $7.88 \pm^{0.00}_{0.00}$ \\
\hline
United Kingdom & $329,828 \pm^{2,345}_{2,401}$ & $41,328 \pm^{49}_{49}$ & $10.74 \pm^{0.12}_{1.54}$ & $0.32 \pm^{0.02}_{0.00}$ & $0.044 \pm^{0.001}_{0.000}$ & $25.12 \pm^{14.74}_{15.13}$ & $6.37 \pm^{0.52}_{0.02}$
\end{tabular}
\end{ruledtabular}
\end{table*}

Other quantities of interest, which are derived from the basic parameters in Table~\ref{tab:1a}, are also shown in Table~\ref{tab:1b}. The first one is the total population number $N$, which within model A is understood to be the total population in susceptible form for the spreading of the disease. We can see that this number is less than the total population in the case of countries that seem to have the epidemics under control, but in some others it can be much larger than the whole world population. This only signals the lack of convergence of model A for the parameters $u_0$ and $\hat{u}_0$, whose fitted values are close to the upper limit in their priors. 

More meaningful is the half-lifetime of the transmission rate represented by $t_{1/2}$, which is directly calculated from parameter $k_1$. The lowest value corresponds to Belgium with the disease decreasing by half every 15.09 days, whereas the highest corresponds to Chile for which the disease decreases by half every 115.51 days. It is interesting to note the relation between $t_{1/2}$ and the measures taken to control the epidemics, as the lower values are for countries with the strictest lockdown measurements.

Almost as meaningful as $t_{1/2}$ are the values of the half-crossing times $t_{50}$ and $t_{D\, 50}$, calculated from Eq.~\eqref{eq:crossing-b}. The lowest values correspond to Belgium, for which the half-crossing occurred just 38 and 35 days after the report of the first deaths. The largest values are for Brazil, resulting in 157 and 124 days, respectively.

\begin{table*}[htp!]
\caption{\label{tab:1b} Fitted values of extra parameters in the case of model A, for the same countries as in Table~\ref{tab:1a}. Shown are the total population number $N$, the half-life time $t_{1/2}$ of the transmission rate $\beta$, see Eq.~\eqref{eq:betatd-b}, the crossing times of half the asymptotic values of cumulative positives and deaths, $t_{50}$ and $t_{D\, 50}$, respectively, see Eqs.~\eqref{eq:crossing}, in number of days after the start of the time series.}
\begin{ruledtabular}
\begin{tabular}{|c|c|c|c|c|}
Country & $N$ & $t_{1/2}$ & $t_{50}$ & $t_{D\,50}$ \\ 
\hline
Mexico & $199,748,406,667,825 \pm^{54,555,581,054,363,484,160}_{199,748,131,793,760}$ & $40.30 \pm^{0.64}_{0.60}$ & $143.81 \pm^{1.60}_{1.52}$ & $134.48 \pm^{0.44}_{0.45}$ \\ 
\hline
Peru & $454,369,877,704,612 \pm^{391,684,701,709,530,300,416}_{454,369,701,437,638}$ & $45.68 \pm^{0.38}_{0.37}$ & $145.80 \pm^{1.06}_{1.04}$ & $157.80 \pm^{0.98}_{0.96}$ \\ 
\hline
Belgium & $150,462 \pm^{61,584}_{73,199}$ & $15.09 \pm^{3.65}_{1.19}$ & $38.43 \pm^{1.51}_{8.32}$ & $35.40 \pm^{0.15}_{0.15}$ \\ 
\hline
Bolivia & $196,415 \pm^{14,886}_{11,032}$ & $81.28 \pm^{8.64}_{7.83}$ & $138.06 \pm^{3.24}_{2.92}$ & $181.65 \pm^{8.53}_{7.61}$ \\ 
\hline
Brazil & $5,743,375,136,031,503 \pm^{388,424,216,445,922,902,016}_{5,743,350,307,608,080}$ & $44.13 \pm^{0.17}_{0.18}$ & $157.58 \pm^{0.60}_{0.62}$ & $124.14 \pm^{0.20}_{0.20}$ \\ 
\hline
Chile & $442,424 \pm^{7,961}_{14,470}$ & $115.51 \pm^{8.39}_{7.65}$ & $92.42 \pm^{1.30}_{2.89}$ & $105.67 \pm^{0.40}_{0.38}$ \\ 
\hline
Ecuador & $2,172,244 \pm^{1,031,514}_{519,254}$ & $37.38 \pm^{0.15}_{0.15}$ & $111.76 \pm^{0.79}_{0.79}$ & $114.32 \pm^{0.53}_{0.54}$ \\ 
\hline
United States & $11,573,101,571,769,328 \pm^{237,514,150,381,025,689,600}_{11,572,844,606,558,488}$ & $31.53 \pm^{0.03}_{0.03}$ & $115.00 \pm^{0.13}_{0.13}$ & $81.67 \pm^{0.07}_{0.07}$ \\ 
\hline
United Kingdom & $18,562,540,734,729 \pm^{47,014,334,860,169,912,320}_{18,562,537,349,892}$ & $15.79 \pm^{0.14}_{0.36}$ & $53.53 \pm^{0.53}_{0.55}$ & $46.80 \pm^{0.09}_{0.09}$
\end{tabular}
\end{ruledtabular}
\end{table*}

\begin{figure}[htp!]
\includegraphics[width=0.49\textwidth]{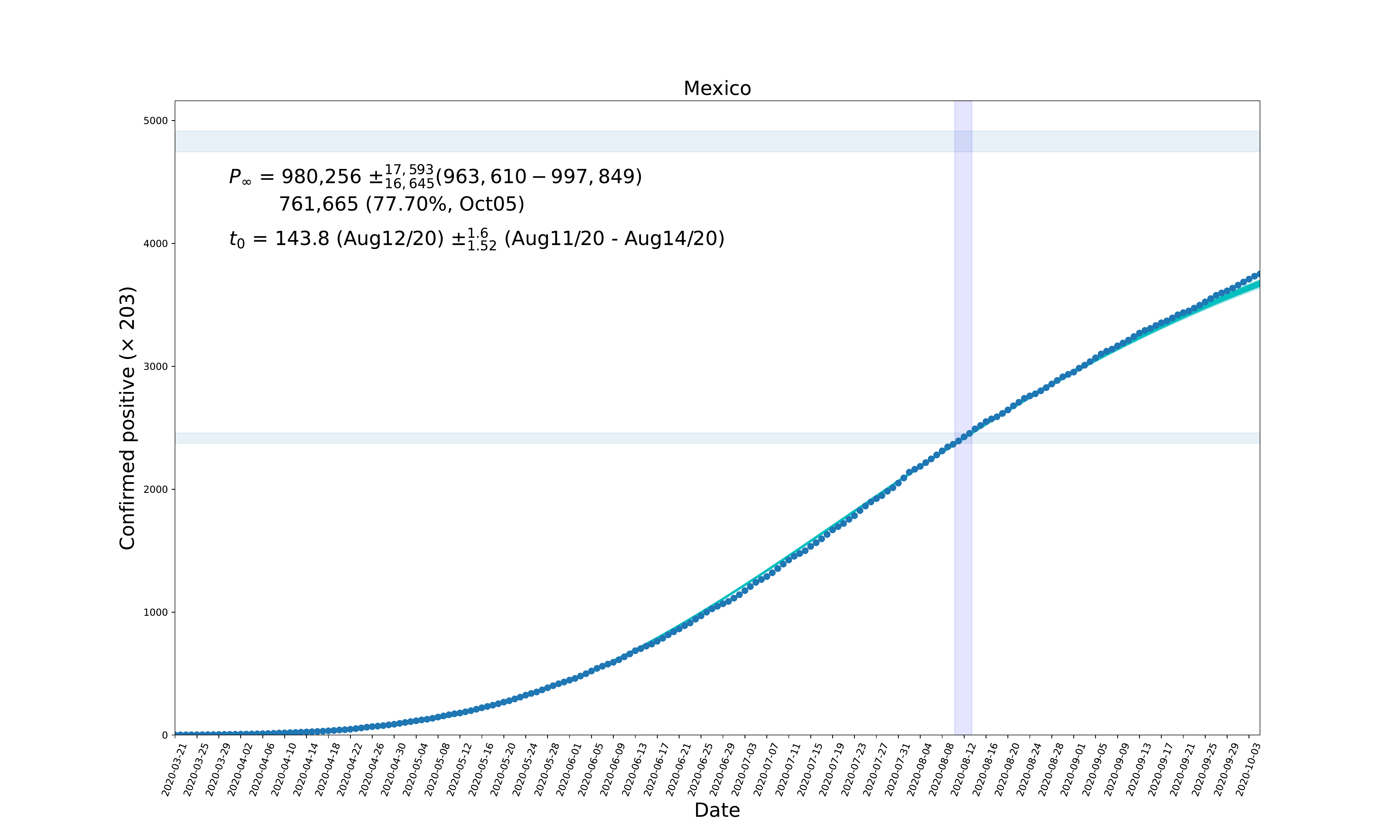}
\includegraphics[width=0.49\textwidth]{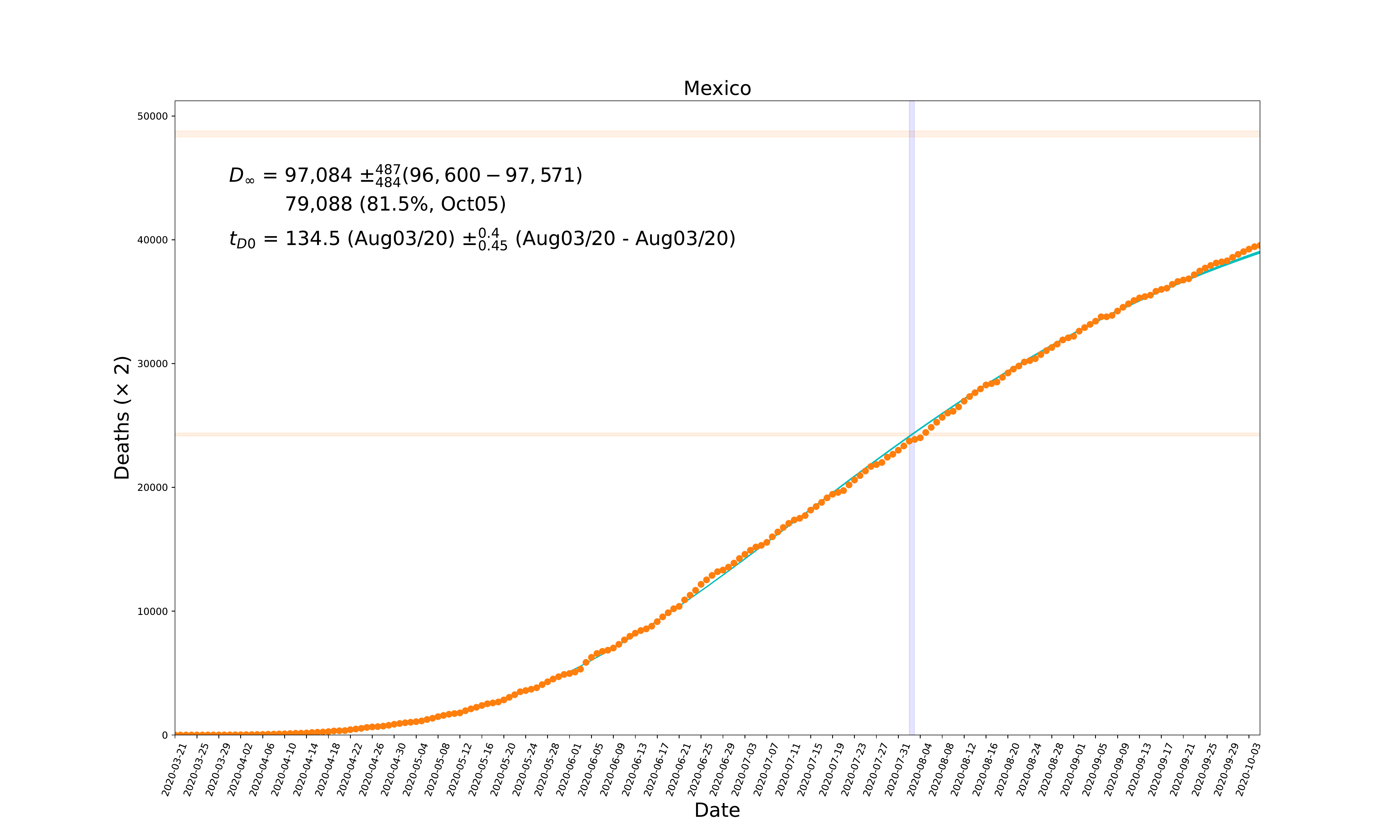}
\includegraphics[width=0.49\textwidth]{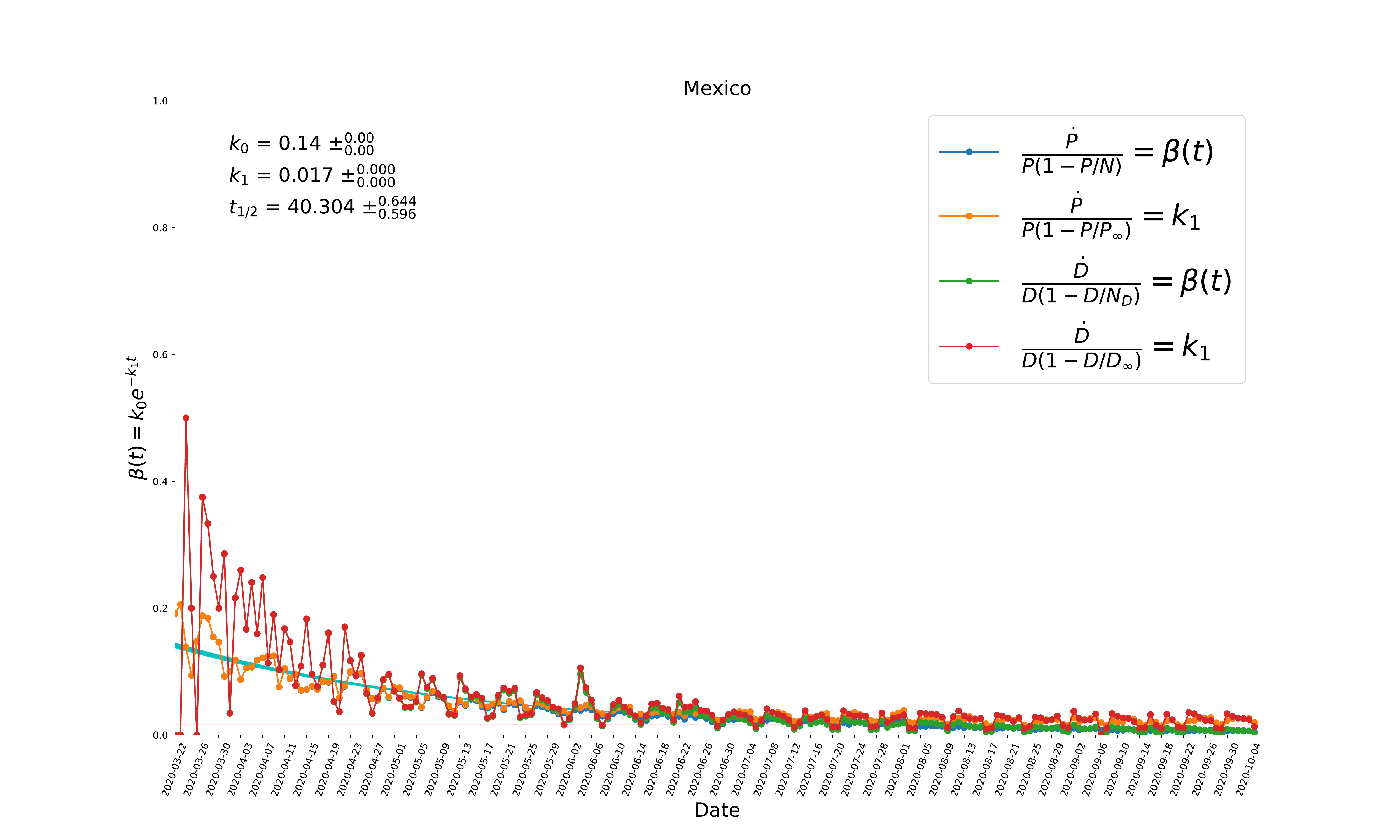}
\caption{\label{fig:modela-b} The resultant evolution curves of accumulated confirmed positives (top panel) and deaths (middle panel) in the case of Mexico, according to the values in the triangle plot in Fig.~\ref{fig:modela-a}. Shown in the figures are the obtained asymptotic values $P_\infty$ and $D_\infty$, see Eq.~\eqref{eq:Iratios} and Table~\ref{tab:1a}, in the top blue-shaded horizontal regions. The vertical blue-shaded regions mark the time of half-crossing in each case; the region for $I_\infty/2$ is also shown for reference. (Bottom panel) The resultant evolution of the transmission rate $\beta$ according to the parametrisation in Eq.~\eqref{eq:betatd-a} and its comparison with data, see Fig.
~\ref{fig:beta} and Tables~\ref{tab:1a} and~\ref{tab:1b}. Shown are also the obtained values of the parameters $k_0$ and $k_1$, and the corresponding half-life time of the disease $t_{1/2}$. The horizontal red-shaded region represents the obtained value of $k_1$, see Eq.~\eqref{eq:limit-k1b}. See the text for more details.}
\end{figure}

In the top and middle panels of Fig.~\ref{fig:modela-b} we show the comparison of data with the estimated evolution curves from our fitting, where the latter are represented by 500 instances of the model using a sample of values around those of maximum likelihood. We see in general a very good agreement with the data, which suggests that Eqs.~\eqref{eq:modela} represent well the evolution of real data. For reference in the plots, we also show the times at half-crossing, around which the number of confirmed positives and deaths were half the asymptotic values $P_\infty$ and $D_\infty$, respectively.

Also, in the bottom panel of Fig.~\ref{fig:modela-b} we show the time evolution of the transmission rate $\beta(t)$ and its comparision with data, represented here by the new daily cases of both confirmed positives and deaths. We see a very good agreement with the combination of data that in principle represents the transmission rate. Likewise, there is good agreement with the combination that represents the parameter $k_1$, which in the plot is represented by the horizontal red-shaded region.

\begin{figure}[htp!]
\includegraphics[width=0.49\textwidth]{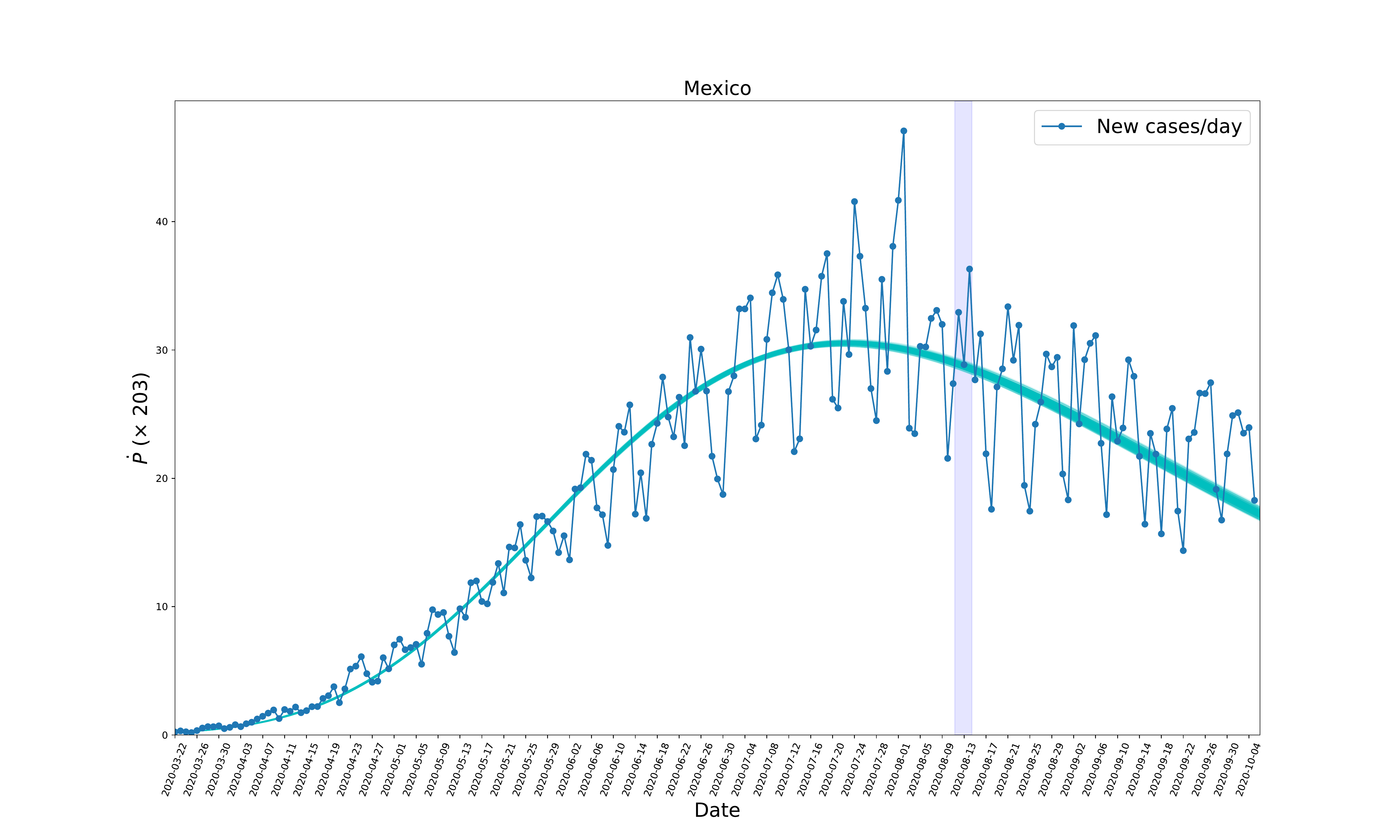}
\includegraphics[width=0.49\textwidth]{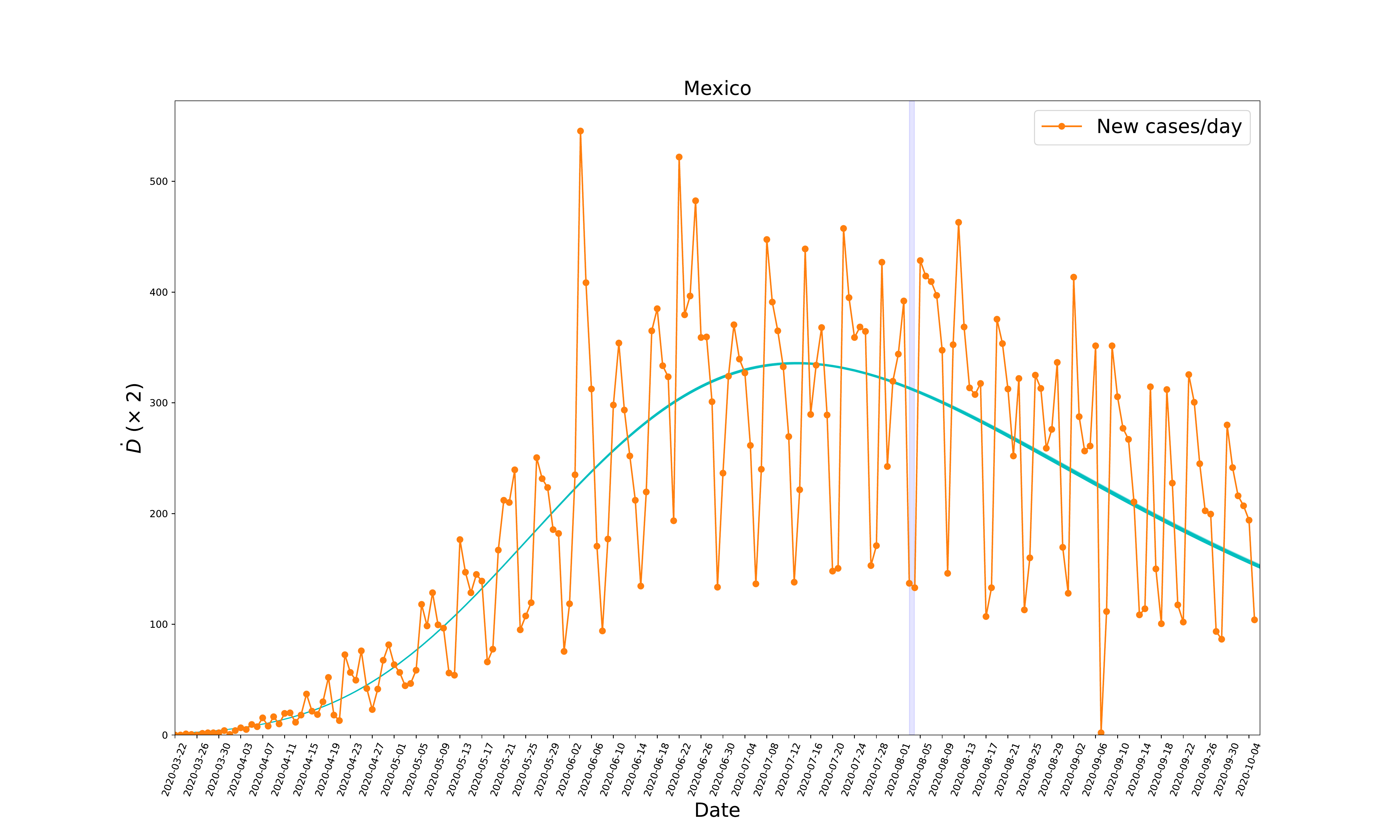}
\caption{\label{fig:modela-c} The derivatives $\dot{P}$ and $\dot{D}$ for positives (top panel) and deaths (bottom panel), respectively, obtained from the data of new daily cases and from the analytical expression~\eqref{eq:si-onedim} using the parameters fitted to the data, see Fig.~\ref{fig:modela-a}. Even though the new daily cases were not used for the fitting, we see a consistent agreement with the results. The blue-shaded vertical regions mark the crossing of half the asymptotic values in each case as in Fig.~\ref{fig:modela-b}. See the text for more details.}
\end{figure}

As an extra comparison with data, we show in Fig.~\ref{fig:modela-c} the derivative of both confirmed positives and deaths as obtained from the fitting to data and using the analytical formula~\eqref{eq:si-onedim} for each case. It must be recalled that the trend of new cases was not used for the fitting to data, and then the aforementioned comparison is useful as an extra validation of the fitting, even if it does not appear to be as good as for the cumulative cases in Fig.~\ref{fig:modela-b}. Additionally, the results show that the derivative of the model can follow the daily evolution of the disease and not just the global trend. 

As a last feature, we show in Fig.~\ref{fig:modela-c} the time of half-crossing with a vertical blue-shaded region, which indicates that the maximum of new daily cases is reached some days before and then the presence of such maximum seems to be a necessary condition for the inflection of the cumulative cases.

\subsection{Model B and fitting to data \label{sec:modelB}}
As explained before, for model B we will consider the following parametrisation of confirmed positives and deaths, in cumulative numbers,
\begin{equation}
     P(t) = P_i \, e^u \, , \quad D(t) = D_i \, e^{u(t-t_D)} \, , \label{eq:modelb}
\end{equation}
where the variable $u$ has the same form as in Eq.~\eqref{eq:betatd-b}. For this parametrisation, we obtain directly that $\dot{P}/P = \beta$ and $\dot{D}/D = \beta e^{k_1 t_D}$, and also obtain the same limit results as in Eqs.~\eqref{eq:limit-k1d}. Explicitly, the functional form of model B~\eqref{eq:modelb} is
\begin{equation}
    F(t) = F_i \exp \left[ \frac{k_0}{k_1} \left( 1 - e^{-k_1 t} \right) \right] \, , \label{eq:gompertz}
\end{equation}
which is no other but the Gompertz function~\cite{gompertz,Winsor1932,Tjorve2017}. In this way one can see that there is a direct connnection between the SI model and the Gompertz function, mediated by an exponentially decaying transmission rate and what we called the large-$N$ limit.

As said before for this simplified model, see Sec.~\ref{sec:large-N} above, one can find analytical expressions for different quantities of interest, which are actually the same one obtains from the Gompertz function~\eqref{eq:gompertz}~\cite{Winsor1932}. The first ones are the asymptotic values at $t \to \infty$, which are given by $P_\infty = P_i e^{k_0/k1}$ and $D_\infty = D_i e^{k_0/k1}$, for confirmed positives and deaths, respectively.

Another analytical result is the time for the inflection of the curve, that we denote by $t_0$ following the nomenklature of the time-independent SI system. The expression is
\begin{subequations}
\begin{equation}
    t_0 = \frac{1}{k_1} \ln \left( \frac{k_0}{k_1} \right) ( +t_D ) \, , \label{eq:inflectionB}
\end{equation}
where we have included the time shift $t_D$ of the function $D(t)$. Likewise, the numbers of confirmed positives and deaths at the inflection time are
\begin{equation}
    P_0 = P_i e^{k_0/k_1 -1} \, , \quad D_0 = D_i e^{k_0/k_1 -1} \, ,
\end{equation}
which is the same functional form for both of them. One can easily see that $P_0 = P_\infty /e$ and $D_0 = D_\infty /e$.

The inflection point also corresponds to a maximum in the derivative of Eqs.~\eqref{eq:modelb}, and then we find
\begin{equation}
    \dot{P}_0 = k_1 P_i e^{k_0/k_1 -1} \, , \quad \dot{D}_0 = k_1 D_i e^{k_0/k_1 -1} \, .
\end{equation}
\end{subequations}

This time the model has five free parameters: $P_i$, $D_i$, $t_D$, $k_0$ and $k_1$, and as in model A the last two of them are common to both confirmed positives and deaths. We will again assume some level of intrinsic error by using a Poisson distribution and the same normalization of the data so that the first point is of order of unity. Given this, we consider flat priors in the form: $0 < P_i < 10$, $0< D_i < 10$, $0 < t_D < 100$ and $0< k_0, \, k_1 < 3$.

\begin{figure}[htp!]
\includegraphics[width=0.49\textwidth]{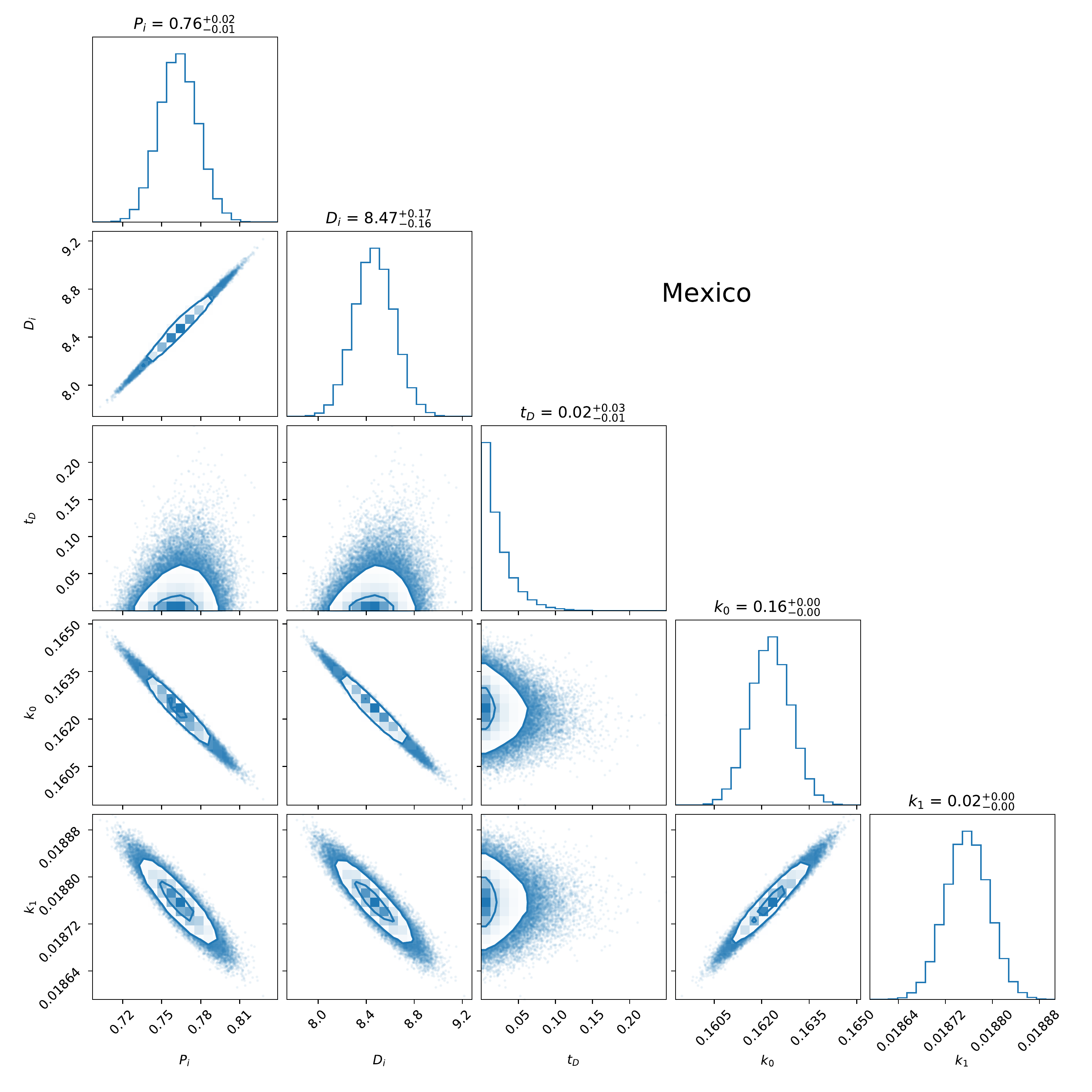}
\caption{\label{fig:modelb-a} Triangle plot of the fitting to data from Mexico of the parameters $P_0$, $D_0$, $t_D$, $k_0$ and $k_1$ of model B described in Sec.~\ref{sec:modelB}. In general, and similarly to model A in Fig.~\ref{fig:modela-a}, all parameters are well constrained by the combination of cumulative confirmed positives and deaths. See the text for more details.}
\end{figure}

We again took the EMCEE algorithm by means of a Python script, using 100 chains with 20,000 steps in each one. The results are shown in Fig.~\ref{fig:modelb-a} for the free parameters of our model. As happened before in model A, the values of $P_0$, $D_0$, $t_0$, $k_0$, and $k_1$, are well constrained by the data and their values are similar and of the same order of magnitude as those of model A. This is seen from a quick comparison of the common parameters in Figs.
~\ref{fig:modela-a} and~\ref{fig:modelb-a}.

The fitting values of the parameters of model B, for the same countries as in model A, are shown in Table~\ref{tab:2a}. Firstly, we notice again that the asymptotic values $P_\infty$ and $D_\infty$ are quite similar to those obtained for model A . This indicates that model B, although simpler than model A, is also a good model to follow the evolution of the cumulative cases. Secondly, the similarity in the results extends to the other common parameters between the models, as is the case of $t_D$, $k_0$ and $k_1$, which again supports the validity of model B to fit the data under simpler assumptions.

\begin{table*}[htp!]
\caption{\label{tab:2a} Fitted values of parameters in model B, see Eqs.~\eqref{eq:modelb}, as obtained from the data of different countries. The confidence regions for the parameters in the case of Mexico are shown in Fig.~\ref{fig:modelb-a}.}
\begin{ruledtabular}
\begin{tabular}{|c|c|c|c|c|c|}
Country & $P_\infty$ & $D_\infty$ & $t_D$ & $k_0$ & $k_1$ (1/days) \\ 
\hline
Mexico & $885,589 \pm^{6,084}_{6,075}$ & $96,848 \pm^{486}_{485}$ & $0.02 \pm^{0.10}_{0.02}$ & $0.16 \pm^{0.00}_{0.00}$ & $0.019 \pm^{0.000}_{0.000}$ \\ 
\hline
Peru & $1,074,659 \pm^{13,147}_{12,693}$ & $51,009 \pm^{530}_{526}$ & $11.98 \pm^{0.58}_{0.58}$ & $0.10 \pm^{0.00}_{0.00}$ & $0.015 \pm^{0.000}_{0.000}$ \\ 
\hline
Belgium & $73,582 \pm^{1,300}_{1,264}$ & $9,748 \pm^{22}_{21}$ & $3.45 \pm^{0.72}_{2.39}$ & $0.55 \pm^{0.12}_{0.03}$ & $0.074 \pm^{0.001}_{0.001}$ \\ 
\hline
Bolivia & $279,451 \pm^{9,928}_{9,223}$ & $16,866 \pm^{674}_{633}$ & $15.14 \pm^{1.13}_{1.13}$ & $0.11 \pm^{0.00}_{0.00}$ & $0.014 \pm^{0.000}_{0.000}$ \\ 
\hline
Brazil & $4,907,718 \pm^{15,697}_{15,750}$ & $169,413 \pm^{387}_{390}$ & $0.00 \pm^{0.00}_{0.00}$ & $0.15 \pm^{0.00}_{0.00}$ & $0.019 \pm^{0.000}_{0.000}$ \\ 
\hline
Chile & $497,634 \pm^{56,953}_{16,172}$ & $14,470 \pm^{89}_{1,571}$ & $14.83 \pm^{0.81}_{14.61}$ & $0.20 \pm^{0.28}_{0.01}$ & $0.025 \pm^{0.007}_{0.000}$ \\ 
\hline
Ecuador & $140,948 \pm^{830}_{857}$ & $10,502 \pm^{58}_{58}$ & $0.05 \pm^{0.23}_{0.05}$ & $0.11 \pm^{0.00}_{0.00}$ & $0.019 \pm^{0.000}_{0.000}$ \\ 
\hline
United States & $5,186,207 \pm^{5,080}_{5,084}$ & $192,021 \pm^{141}_{142}$ & $0.00 \pm^{0.00}_{0.00}$ & $0.23 \pm^{0.00}_{0.00}$ & $0.028 \pm^{0.000}_{0.000}$ \\ 
\hline
United Kingdom & $312,737 \pm^{2,051}_{2,307}$ & $41,127 \pm^{50}_{49}$ & $0.95 \pm^{1.21}_{0.36}$ & $0.66 \pm^{0.02}_{0.06}$ & $0.060 \pm^{0.000}_{0.000}$
\end{tabular}
\end{ruledtabular}
\end{table*}

The same happens when one compares the fitting results of the half-lifetime $t_{1/2}$ with those in Table~\ref{tab:1b}, they are very similar one to each other for the respective countries. The similarity extends for the case of the inflection times $t_0$ and $t_{D0}$, which are lower than the half-crossing times of model A. This is as expected, given that in model A the half-crossing should happen after the inflection of the resultant evolution curve. As in Table~\ref{tab:1b}, we find that the lowest characteristic times correspond to Belgium, whereas the highest correspond to Bolivia.

\begin{table}[htp!]
\caption{\label{tab:2b} Fitted values of extra parameters in the case of model B, for the same countries as in Table~\ref{tab:1a}. Shown are the half-life time $t_{1/2}$ of the transmission rate $\beta$, see Eq.~\eqref{eq:betatd-b}, and the inflection times of cumulative positives and deaths, $t_0$ and $t_{D0}$, respectively, see Eqs.~\eqref{eq:inflectionB}}
\begin{ruledtabular}
\begin{tabular}{|c|c|c|c|}
Country & $t_{1/2}$ & $t_0$ & $t_{D0}$ \\ 
\hline
Mexico & $36.95 \pm^{0.19}_{0.19}$ & $115.03 \pm^{0.35}_{0.36}$ & $115.06 \pm^{0.35}_{0.35}$ \\ 
\hline
Peru & $45.63 \pm^{0.37}_{0.36}$ & $121.59 \pm^{0.89}_{0.87}$ & $133.56 \pm^{0.79}_{0.78}$ \\ 
\hline
Belgium & $9.40 \pm^{0.09}_{0.10}$ & $27.38 \pm^{2.40}_{0.71}$ & $30.83 \pm^{0.12}_{0.13}$ \\ 
\hline
Bolivia & $50.31 \pm^{1.08}_{1.03}$ & $147.92 \pm^{2.54}_{2.40}$ & $163.06 \pm^{2.73}_{2.61}$ \\ 
\hline
Brazil & $36.12 \pm^{0.09}_{0.09}$ & $107.45 \pm^{0.16}_{0.17}$ & $107.45 \pm^{0.16}_{0.16}$ \\ 
\hline
Chile & $28.18 \pm^{0.22}_{6.55}$ & $86.18 \pm^{10.85}_{1.75}$ & $100.99 \pm^{0.38}_{6.52}$ \\ 
\hline
Ecuador & $37.30 \pm^{0.16}_{0.17}$ & $93.90 \pm^{0.44}_{0.46}$ & $93.96 \pm^{0.42}_{0.42}$ \\ 
\hline
United States & $25.04 \pm^{0.02}_{0.02}$ & $76.14 \pm^{0.05}_{0.05}$ & $76.14 \pm^{0.05}_{0.05}$ \\ 
\hline
United Kingdom & $11.59 \pm^{0.08}_{0.05}$ & $40.24 \pm^{0.38}_{1.16}$ & $41.19 \pm^{0.07}_{0.07}$
\end{tabular}
\end{ruledtabular}
\end{table}

\begin{figure}[htp!]
\includegraphics[width=0.49\textwidth]{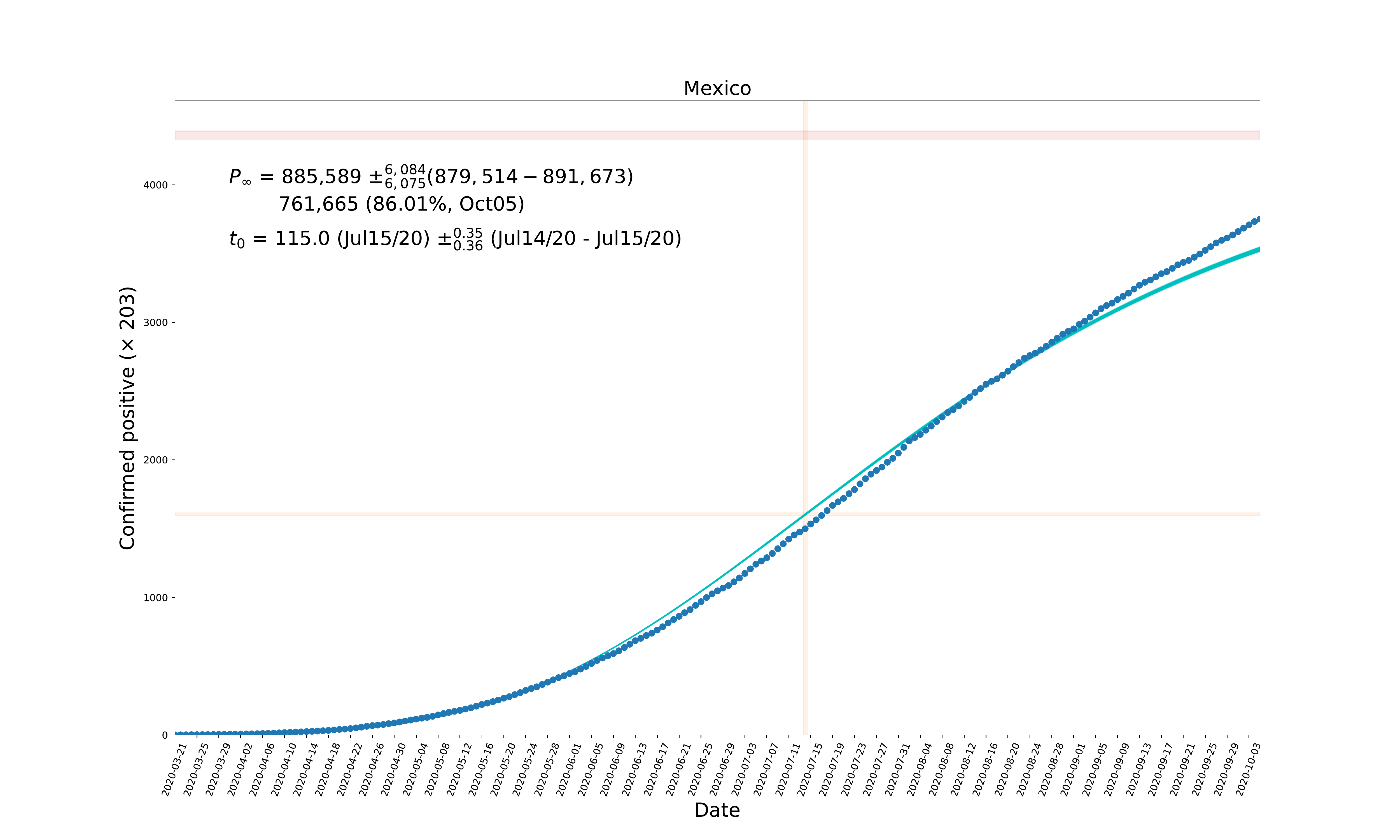}
\includegraphics[width=0.49\textwidth]{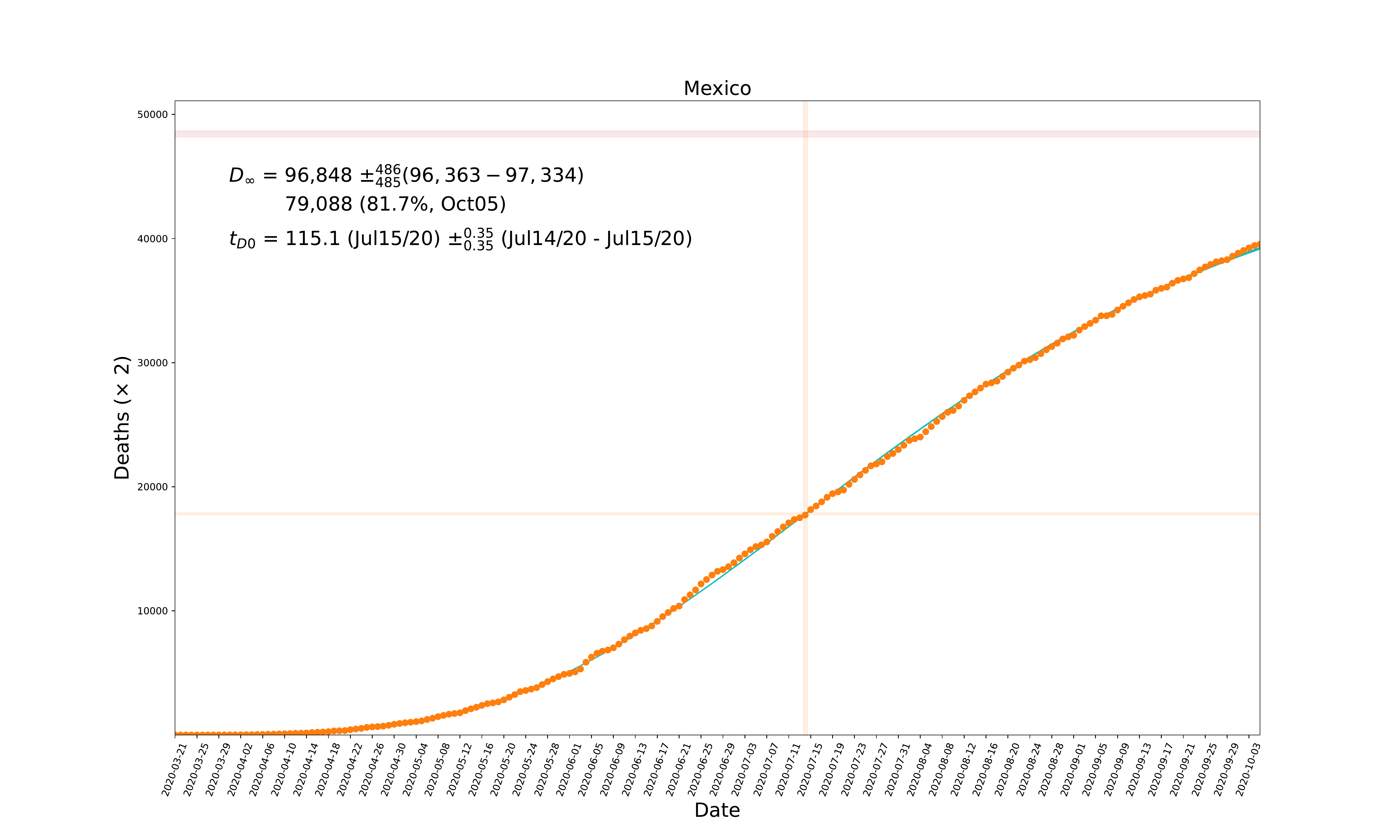}
\includegraphics[width=0.49\textwidth]{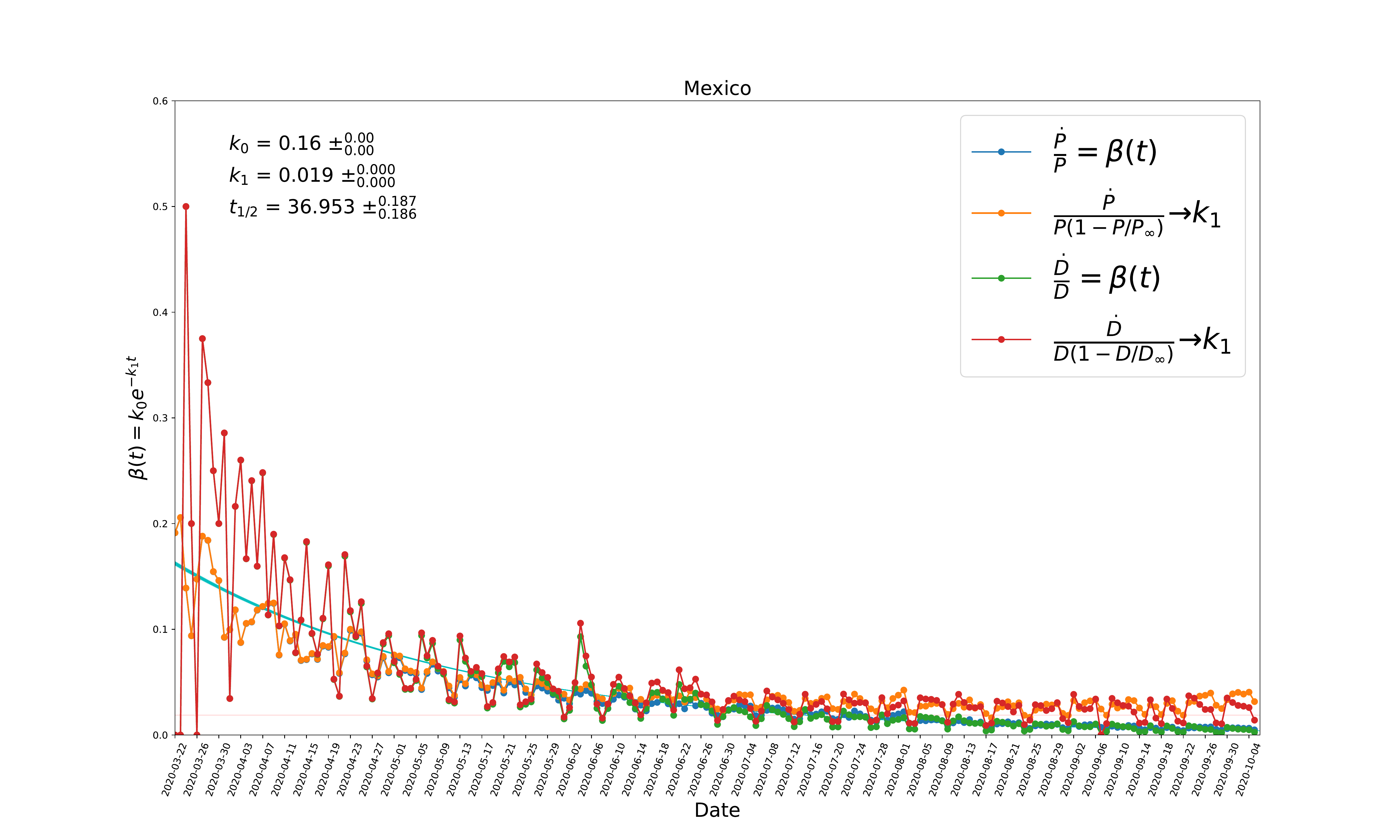}
\caption{\label{fig:modelb-b} The resultant evolution curves of cumulative confirmed positives (top panel) and deaths (middle panel) for model B in the case of Mexico, according to the values in the triangle plot in Fig.~\ref{fig:modelb-a}. Shown in the figures are the obtained asymptotic values $P_\infty$ and $D_\infty$, see Eq.~\eqref{eq:Iratios} and Table~\ref{tab:2b}, in the top blue-shaded horizontal regions in each plot. The vertical blue-shaded regions mark the time inflection in each case; the region for $P(t_0)$ is also shown for reference. (Bottom panel) The resultant evolution of the transmission rate $\beta$ according to the parametrisation in Eq.~\eqref{eq:betatd-a} and its comparison with data, see Fig.~\ref{fig:beta} and Tables~\ref{tab:2a} and~\ref{tab:2b}. Shown are also the obtained values of the parameters $k_0$ and $k_1$, and the corresponding half-life time of the disease $t_{1/2}$. The horizontal red-shaded region represents the obtained value of $k_1$, see Eq.~\eqref{eq:limit-k1b}. See the text for more details.}
\end{figure}

We repeated the comparisons between the data and model B in the same form as in Fig.~\ref{fig:modela-b}, the new results are shown in Fig.~\ref{fig:modelb-b}. As anticipated in Sec.~\ref{sec:time-dependent-data}, model B is also good at following the trend of the data and the only changes are in values of the fitted parameters, which also result in changes of the final quantities $P_\infty$ and $D_\infty$, although the obtained values are consistent one to each other in their order of magnitude in the two models. 

\begin{figure}[htp!]
\includegraphics[width=0.49\textwidth]{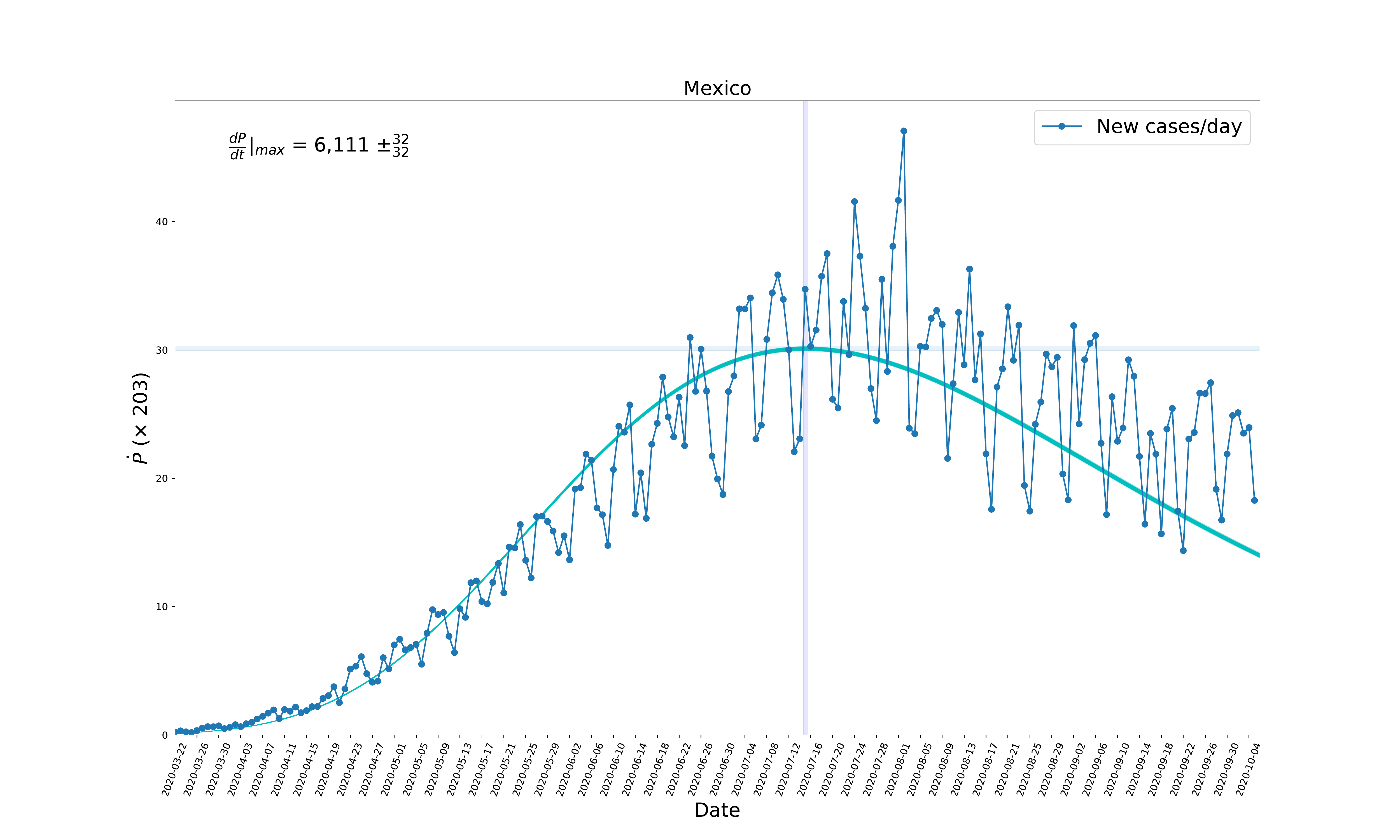}
\includegraphics[width=0.49\textwidth]{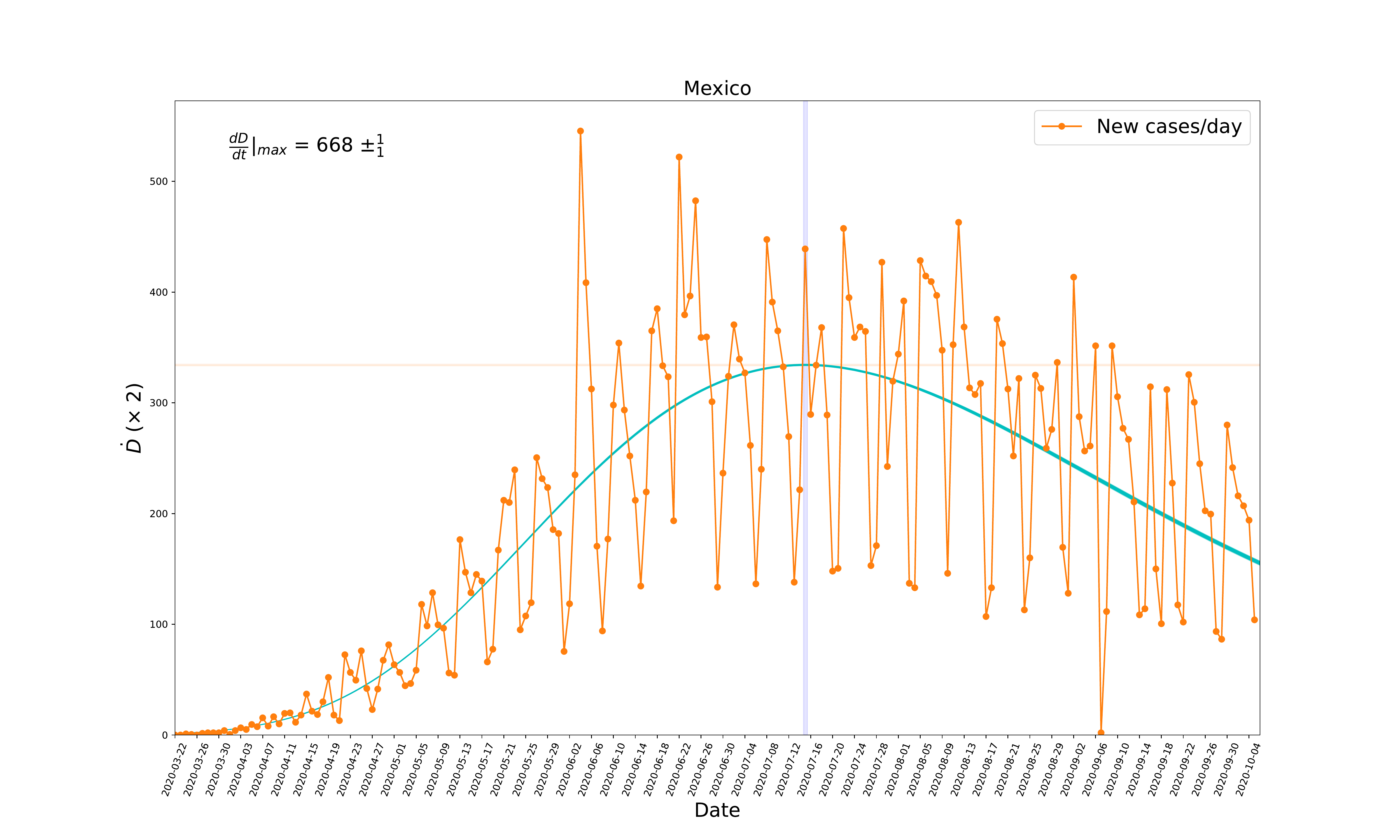}
\caption{\label{fig:modelb-c} The derivatives $\dot{P}$ and $\dot{D}$ for positives (top panel) and deaths (bottom panel), respectively, obtained from the data of new daily cases and from the analytical expression of model B, see Eq.~\eqref{eq:modelb}, with the parameters fitted to the data, see Fig.~\ref{fig:modelb-a}. Even though the new daily cases were not used for the fitting, we see a consistent agreement with the results. The blue-shaded vertical regions mark the inflection time in each case as in Fig.~\ref{fig:modelb-b}, whereas the blue-shaded horizontal regions mark the maximum values of each case. See the text for more details.}
\end{figure}

Finally, in Fig.~\ref{fig:modelb-c} we show the comparison of the derivatives of model B for both the confirmed positives and deaths with their corresponding data. We also see a good agreement in both the time profile and the location and height of the maximum points in the two plots. This time model B is more easily manageable, and we show the maximum daily new cases as expected from the theoretical expectations. Notice that the time at the location of the maximum is the same as that of the inflection time $t_0$ in the evolution of the cumulative cases, see Fig.~\ref{fig:modelb-b}.

\section{Final comments \label{sec:final}}
We have presented a generalization of the well-known SI system to include the possibility of a time-dependent transmission rate, and used a particular exponential-like parametrisation of it to fit and describe the evolution of real data for the epidemics of Covid-19.

Although the use of time-dependent transmission rates is not new in the literature of epidemic models, this is the first time that such approach have been applied to the SI system, for which there exists an analytical form of the general solution. The latter is the standard logistic function but now with a generalized time variable that results from the direct integration of the transmission rate. These features simplifies the handling of the model and eases its comparison with data.

For the functional form of the time-dependent transmission rate we chose a decaying exponential with two free parameters, the first one for the initial value of the transmission rate and the second one for its decay rate. In this form, our generalized model has the enough complexity to fit the data reliably, whereas at the same time provides meaningful quantities to describe the evolution of cumulative positives and deaths.

One of those meaningful quantities is the decay rate, and the related one the half-lifetime of the transmission rate. It was clear from our results that the half-lifetime is shorter for countries that have taken the strictest measures of public containment. Other countries seem to have an almost three times larger half-lifetime, which means that it will take longer for them to tame the epidemics. Even though in medical terms one may wish to have a slow epidemics to avoid saturation of hospital services, this also means that containment measures will have to take place for longer too, which may result in a general public tired of the governmental intervention.

Our model also suggest an interesting correlation between the initial value of the transmission rate and its decay rate: countries that experienced a faster dissemination at the beginning are the ones that report a larger decay rate, as is the case of Belgium and the United Kingdom. Likewise, countries with a slow initial spreading seem to be the ones with also a lower decay rate. These last countries were not hit as badly as others at the start of their epidemics, but all so far indicates, according to our models, that they will have some of the highest death tolls in the world.

A surprising result was the connection of the SI system with the Gompertz function. As explained in the text, such connection requires some assumptions in between, mainly that the transmission rate is time-dependent with an exponential form, and that we consider a large susceptible population. Although the Gompertz function is quite useful for a plethora of growth phenomena, ours is the first study that shows a derivation of it from a infectious model.

One final note on the fittings we obtained for the models. The Bayesian inference is an appropriate method to fit data if one faces a unique realization of the natural phenomenon under study, which is the case of the present epidemics of Covid-19, as the data reported by each country is not at al the result of repeated experiments under controled conditions. In this respect, the Bayesian analysis allows us to do a sampling of values of the free parameters around the point of maximum likelihood. This does not mean that one finds the best and only fitting to data, but the best possible fit given the proposed model. This helps to explain the well defined confidence regions in the triangle plots of the fitted parameters, even though the resultant curves may not even look good by eye when compared to data (see for instance model C in the Appendix~\ref{sec:modelC}).

The numbers reported here are not definitive, and they may change considerably if the epidemics follow a different evolution in the near future. However, we believe that our models may be helpful in characterizing the present evolution of the disease and can be taken in consideration to decide about further public measures to handle the epidemics.

\begin{acknowledgments}
We are thankful to Argelia Bernal, Juan Barranco, Nana Cabo, Gustavo Niz and Demi\'an Mayorga for enlightening conversations and comments. AXG-M acknowledges support from C\'atedras CONACYT. This work was partially supported by Programa para el Desarrollo Profesional Docente; Direcci\'on de Apoyo a la Investigaci\'on y al Posgrado, Universidad de Guanajuato, research Grant No. 036/2020; CONACyT M\'exico under Grants No. A1-S-17899, 286897, 297771; and the Instituto Avanzado de Cosmolog\'ia Collaboration.
\end{acknowledgments}

\appendix
\section{Parametric form of the SI system}
For completeness, we also show here an alternative derivation of the solution of the SI system in a parametric form. Considering a change of time variable by the relation $d \tau = (I/N) dt$, the solution is
\begin{subequations}
\label{eq:si-param}
\begin{equation}
    S(\tau) = S_i e^{-\beta \tau} \, , \quad I(\tau) = N - S(\tau) \, , \label{eq:si-param-a}
\end{equation}
where the explicit relation between $t$ and $\tau$ is
\begin{equation}
    t = \int^\tau_0 \frac{d\tau}{1 - S(\tau)/N} \, . \label{eq:si-param-b}
\end{equation}
\end{subequations}

The above solution can also be generalized in the case of a time-dependent transmission rate $\beta = \beta(\tau)$, and the only change is the solution for the susceptibles,
\begin{equation}
    S(\tau) = S_i \exp \left[ -\int^\tau_0 \beta(x) dx \right] \, .
\end{equation}
The solutions for the number of infectives $I(\tau)$ and the relation between the time variables $t$ and $\tau$ is the same as in Eqs.~\eqref{eq:si-param}.

\section{Time-dependent SI model}
Here we present an alternative method to solve the time-dependent SI system~\eqref{eq:si-general}. If we take a general function $\xi(t) = \beta S/N$, then $\dot{I}=\xi I$ and the number of infecteds is given by
\begin{subequations}
\label{eq:si-general}
\begin{equation}
    I(t) = I_i \exp \left[ \int^t_0 \xi(x) \, dx \right] \, , \label{eq:si-general-a}
\end{equation}
whereas that of susceptibles, from the direct integration of the equation $\dot{S} = - \xi(t) I(t)$, is
\begin{equation}
    S(t) = S_i + I_i - I(t) \, . \label{eq:si-general-b}
\end{equation}
\end{subequations}
Equation~\eqref{eq:si-general-b} clearly shows the consistency of the solution as from it one recovers the constraint equation $S+I = S_i +I_i = N$.

For the case in which $\beta = \mathrm{const.}$, the evolution equation for the susceptibles becomes a constraint equation for the functional form of $\xi (t)$, namely,
\begin{equation}
    \xi (t) = \beta \left[ 1 - \frac{I_i}{N} \exp \left( \int^t_0 \xi(x) \, dx \right) \right] \, , \label{eq:si-constraint}
\end{equation}
The exact solution of Eq.~\eqref{eq:si-constraint} exists, and can be obtained from the exact solution of the SI system, see Eq.~\eqref{eq:si-onedim}. Hence, we find from $\xi(t) = \dot{I}/I$, or from a direct substitution in Eq.~\eqref{eq:si-constraint}, that
\begin{equation}
    \xi (t) = \beta \left[ 1+e^{\beta(t-t_0)} \right]^{-1} \, .
\end{equation}

Notice that $\xi(t \to -\infty) = -\beta$ and $\xi(t \to \infty) = 0$, and in this sense $\xi(t)$ can be considered a kind of time-dependent transmission rate, in which the time dependence is clearly inherited from the evolution of the factor $S(t)/N$. Actually, the solutions obtained from Eqs.~\eqref{eq:si-general} are
\begin{equation}
    \frac{S(t)}{N} = \left[ 1+e^{\beta(t-t_0)} \right]^{-1} \, , \quad I(t) = \frac{I_i \left( 1+e^{\beta t_0} \right)}{1+e^{-\beta(t-t_0)}} \, ,
\end{equation}
and then one can also see, after a quick comparison with Eq.~\eqref{eq:logistic}, that $N=I_i (1+e^{\beta t_0})$.

The inflection point $t_\ast$ in the evolution of infectives is found from the condition $\ddot{I}(t_\ast)=0$, which explicitly reads
\begin{equation}
    \left( \dot{\xi} + \xi^2 \right)_{t_\ast} = 0 \, .
\end{equation}
We can see that a necessary condition to satisfy the foregoing equation is $\dot{\xi} <0$, ie, that $\xi$ is a decaying function of time. 

Another property of the general solution~\eqref{eq:si-general-a} is that $N$, the total population number, is a free parameter. This is quite convenient, as it means that
\begin{equation}
    \lim_{N \to \infty} \frac{I(t)}{N} = 0 \, ,
\end{equation}
except for the case $\beta = \mathrm{const.}$, for which
\begin{equation}
     \lim_{N \to \infty} \frac{I(t)}{N} =  \left[ 1+e^{\beta(t_0-t)} \right]^{-1} \, .
\end{equation}
The reason behind these results is the mechanism that puts an end to the epidemics. In the case $\beta = \mathrm{const.}$ the epidemics ends because of the exhaustion of the susceptible population, whereas in any other time-dependent case it ends because the transmission rate decays. In the first case we then need to know the total population beforehand, whereas in the second one that number is irrelevant as long as $I(t) < N$. 

Moreover, we can calculate the equivalent time-dependent transmission rate as
\begin{equation}
    \xi(t) = \beta(t) S(t)/N =  \beta(t) (1-I(t)/N) \, , 
\end{equation}
and then $\xi(t) \to \beta(t)$ in the limit $N \to \infty$. Thus, the solution of the infected population in the same limit can simply be written as
\begin{equation}
    I(t) = I_i \exp \left[ \int^t_0 \beta(x) \, dx \right] \, ,
\end{equation}
which is the same functional form as model B in Sec.~\ref{sec:modelB}, see also Eq.~\eqref{eq:degeneracy2-limit}.

\section{Fitting to data with the time-independent SI system \label{sec:modelC}}
For completeness, we show in this appendix the fitting to data using the time-independent logistic function~\eqref{eq:logistic}. We use the data from Mexico, for a easy comparison with the results of model A and B in the main text.

The explicit functions for cumulative positives and deaths are
\begin{equation}
     P(t) = P_0 \, \frac{1+e^{k_0 t_P}}{1+e^{k_0(t_P-t)}} \, , \quad D(t) = D_0 \, \frac{1+e^{k_0 t_D}}{1+e^{k_0(t_D-t)}} \, . 
\end{equation}

\begin{figure*}[htp!]
\includegraphics[width=0.49\textwidth]{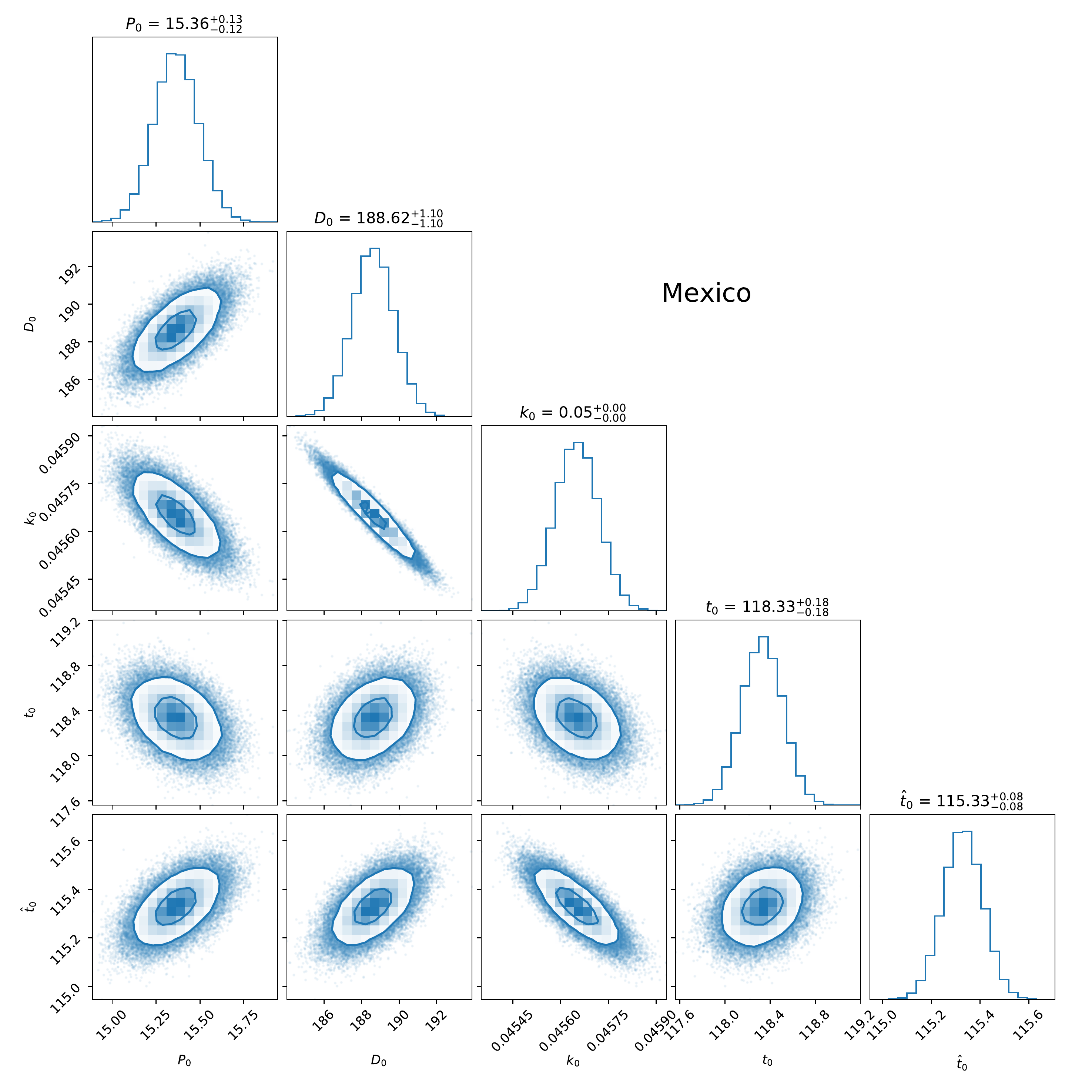}
\includegraphics[width=0.49\textwidth]{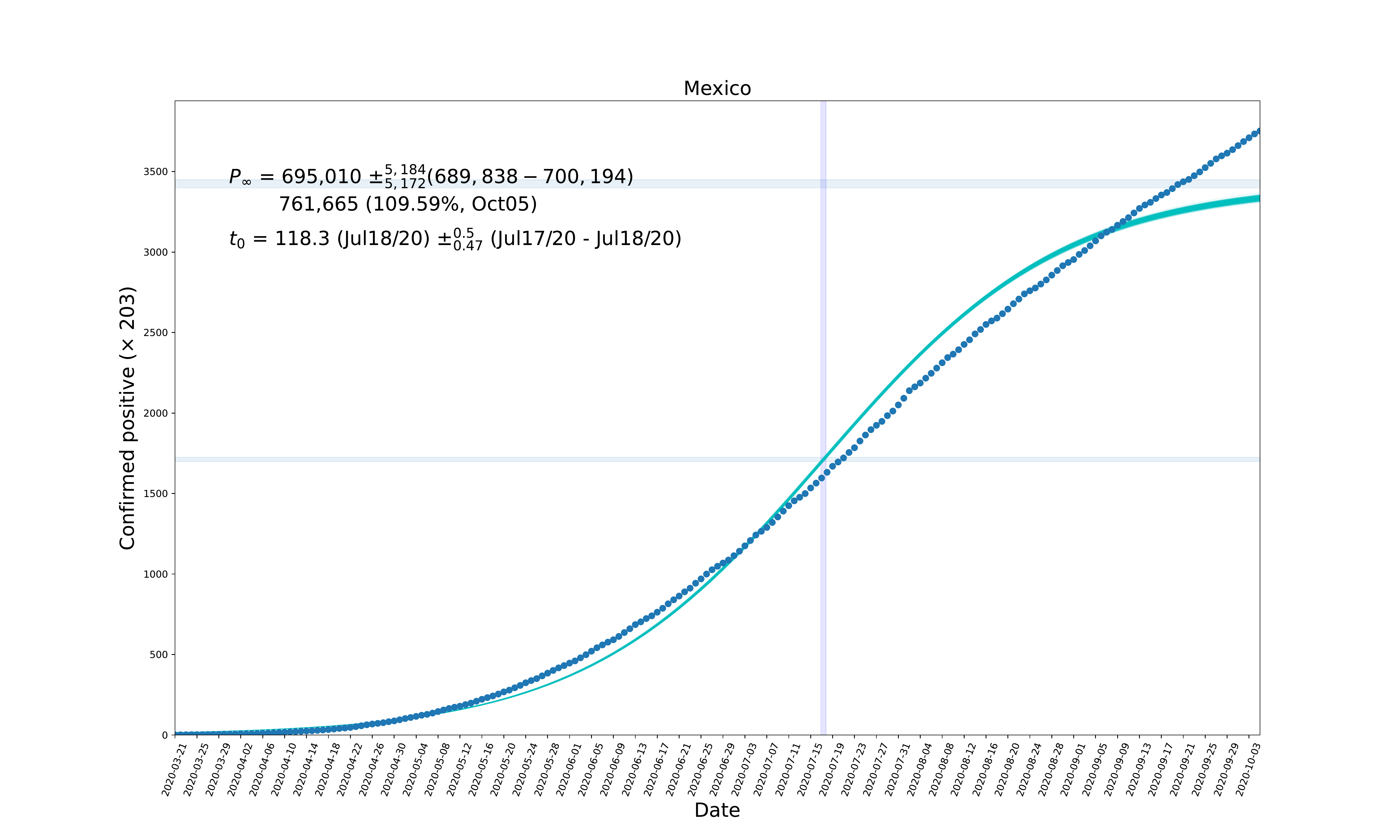}
\includegraphics[width=0.49\textwidth]{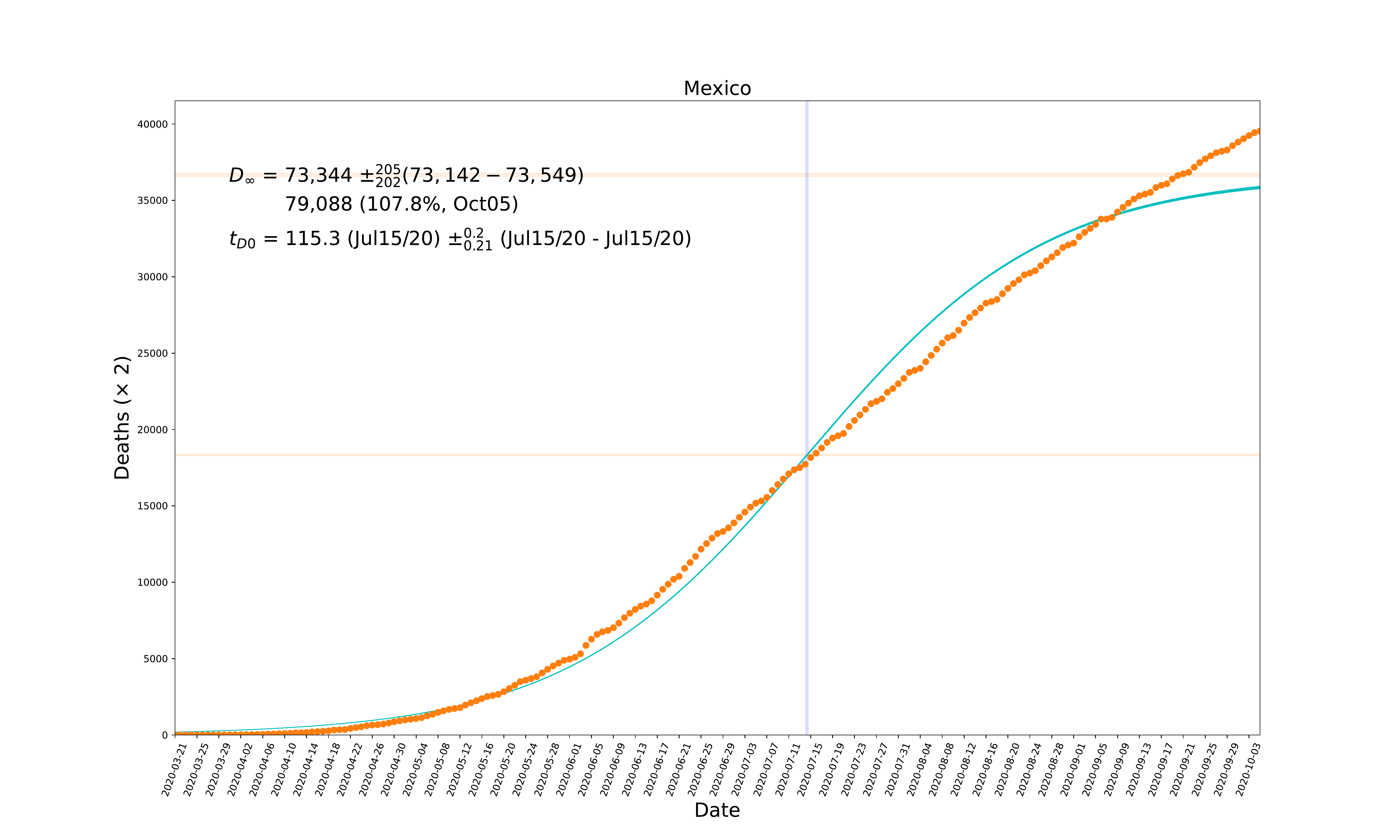}
\includegraphics[width=0.49\textwidth]{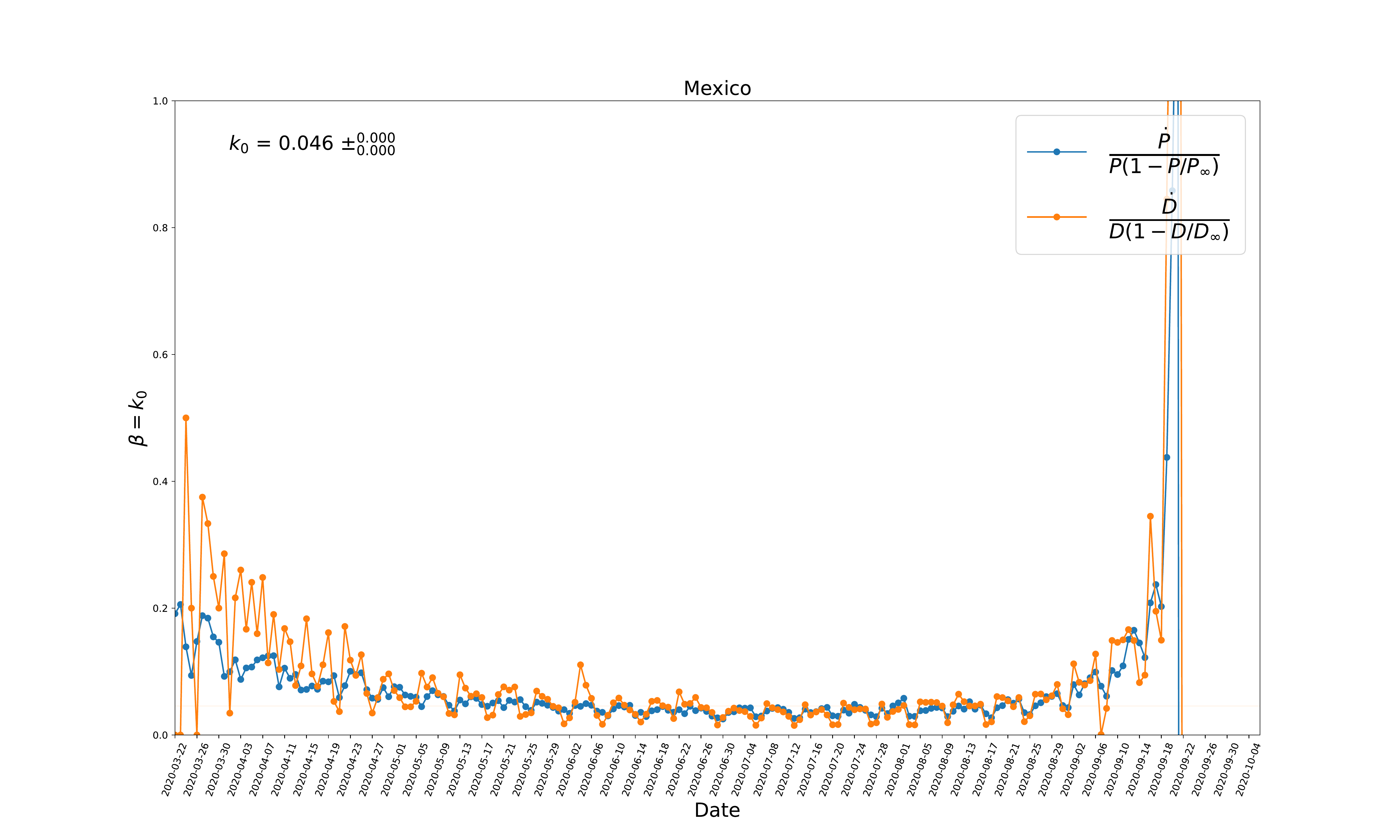}
\includegraphics[width=0.49\textwidth]{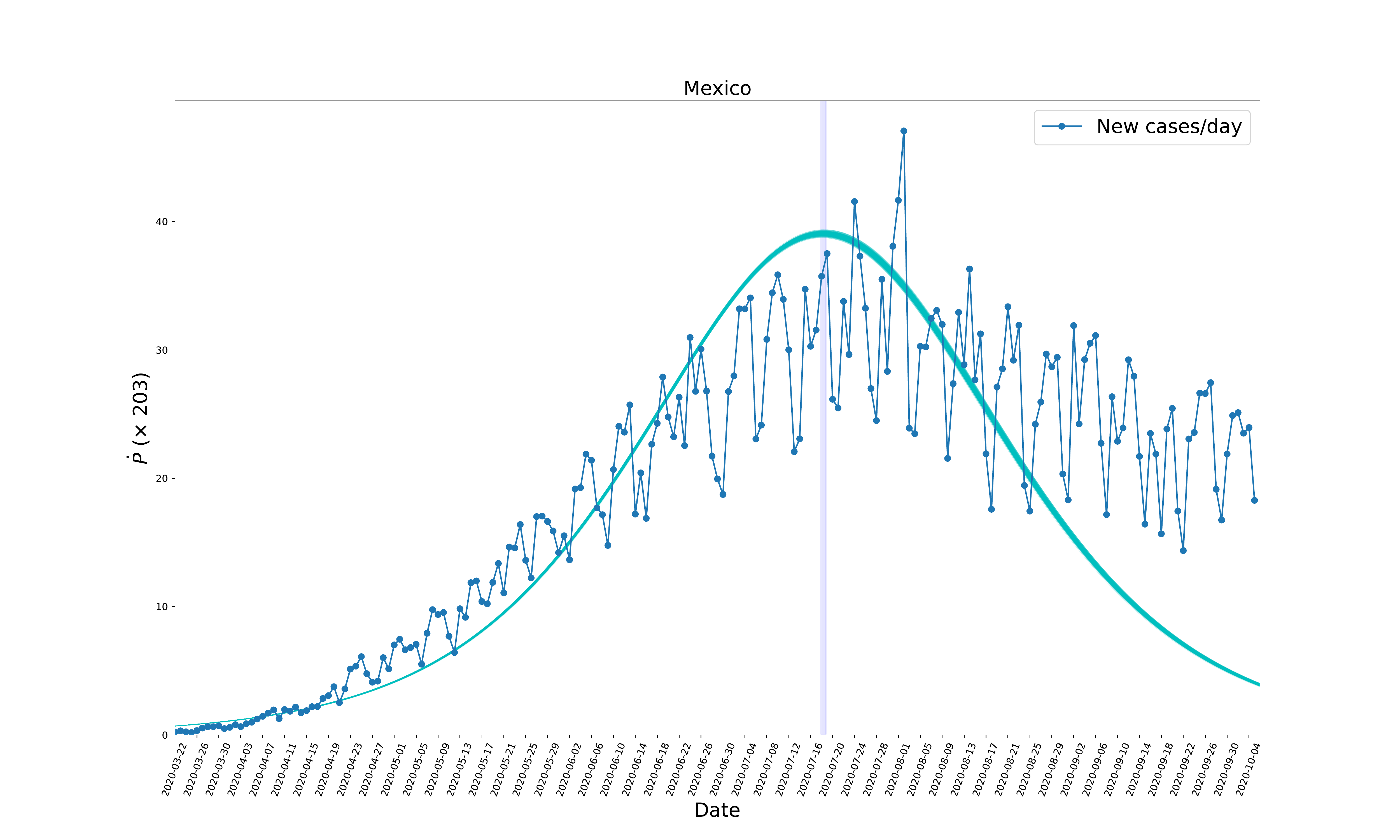}
\includegraphics[width=0.49\textwidth]{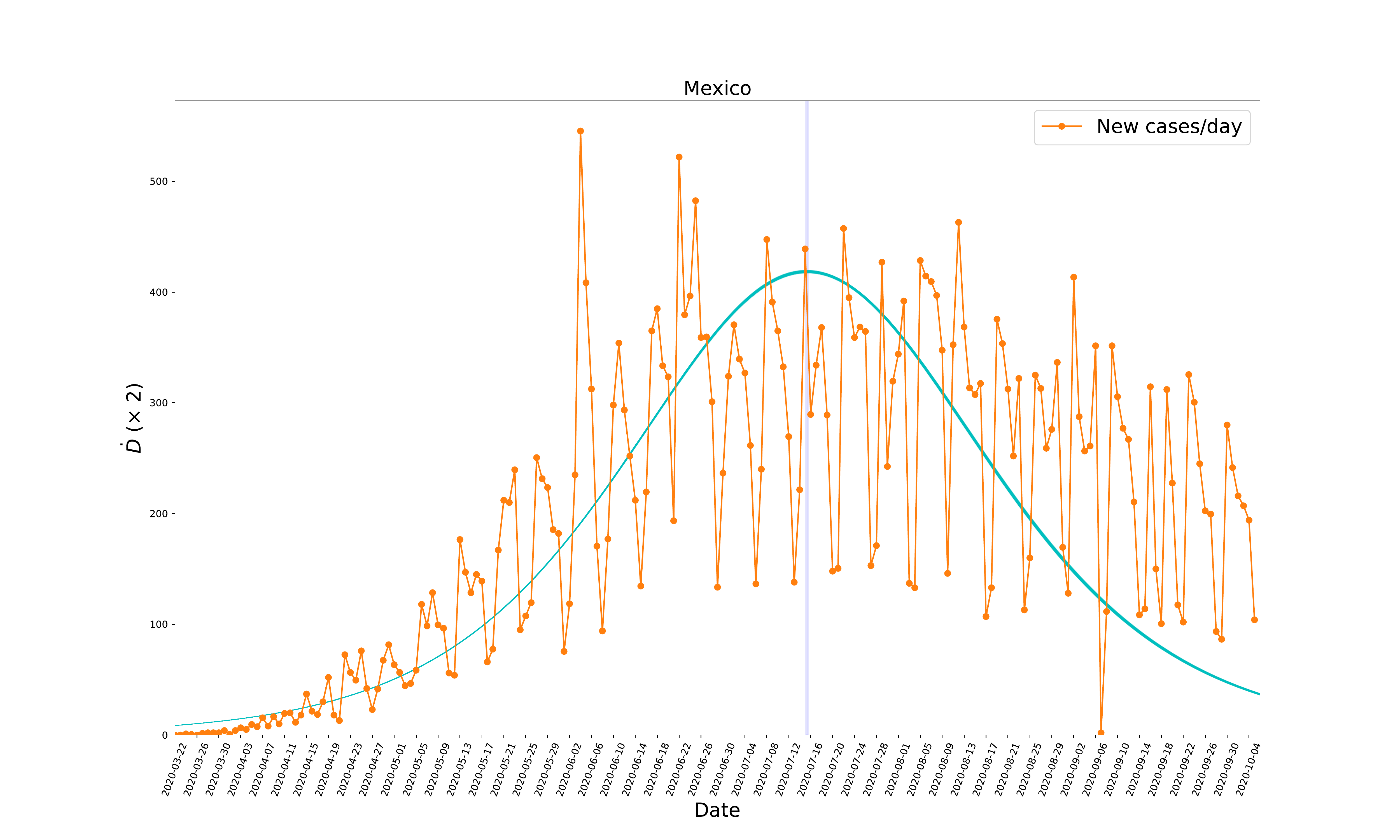}
\caption{\label{fig:modelc-b} (Top left panel) Triangle plot of the fitting to data from Mexico of the parameters $P_0$, $D_0$, $t_P$ $t_D$ and $k_0$ of model C. All parameters are well constrained by the combination of accumulated confirmed positives and deaths. The resultant evolution curves of cumulative confirmed positives (top right panel) and deaths (middle left panel), according to model C with the values of its parameters as shown in the triangle plot. Shown in the figures are the obtained asymptotic values $P_\infty$ and $D_\infty$ in the top blue-shaded horizontal regions. The vertical blue-shaded regions mark the inflection time in each case, $t_P$ and $t_D$, respectively; the region for $P(t_P)$ and $D(t_D)$ are also shown for reference. (Middle right panel) The resultant transmission rate $\beta$ and its comparison with data. The horizontal red-shaded region represents the obtained value of $k_0$. (Bottom) The derivatives $\dot{P}$ and $\dot{D}$ for positives (left panel) and deaths (right panel), respectively, obtained from the data of new daily cases and from the analytical expression~\eqref{eq:si-onedim} using the parameters fitted in the triangle plot. The blue-shaded vertical regions mark the inflection time in each case as in Fig.~\ref{fig:modelc-b}. See the text for more details.}
\end{figure*}

Figure~\ref{fig:modelc-b} is a compilation of all results related to model C. The triangle plot (top left panel) shows the confidence regions for the free parameters in the model, namely $P_0$, $D_0$, $t_0$ and $t_{D0}$. All parameters are well constrained and their confidence regions are well defined too. However, when model C itself is compared directly with data, in the top right and middle left panels of Fig.~\ref{fig:modelc-b}, we see that the agreement is not good enough, as in both cases of cumulative positives and deaths the asymptotic values are lower than the value of the last data point. 

Likewise, in the comparison with new daily cases, in the middle right panel of Fig.~\ref{fig:modelc-b}, we see that the fit is not good either, which confirms again that a constant transmission rate is not appropriate for the complexity of the data.

Finally, in the bottom panels of Fig.~\ref{fig:modelc-b} we also compare the derivative of model C for a direct comparison with new daily cases. Here we see a clear offset of the predicted maximum with respect to that of the data, which leads us to conclude, with all results together, that the simple logistic function~\eqref{eq:logistic} is not appropriate to describe the evolution of real data from the epidemics of Covid-19. 

\bibliography{sorsamp}

\end{document}